\documentclass[reprint,superscriptaddress,nofootinbib,amsmath,amssymb,aps,prd,nobalancelastpage]{revtex4-2}
\usepackage[
  letterpaper,
  left=1.3cm,
  right=1.3cm,
  top=1.4cm,
  bottom=1.6cm
]{geometry}

\usepackage{soul}
\usepackage{etoolbox}

\makeatletter
\def\frontmatter@finalspace{\vspace*{3pt}}%
\makeatother

\usepackage{float}
\usepackage{booktabs}
\usepackage{comment}
\usepackage{graphicx}
\usepackage{dcolumn}
\usepackage{bm}
\usepackage[dvipsnames]{xcolor}
\usepackage{amsmath}
\usepackage{tabularx,array}
\usepackage[T1]{fontenc}
\usepackage[utf8]{inputenc}
\usepackage[english]{babel}
\usepackage{lmodern}
\usepackage[normalem]{ulem}

\usepackage[colorlinks,bookmarks]{hyperref}
\definecolor{linkblue}{rgb}{0,0,0.8}
\definecolor{linkgreen}{rgb}{0,0.5,0}
\hypersetup{linkcolor=linkblue, citecolor=linkgreen, urlcolor=linkblue}

\definecolor{valecol}{rgb}{0,0.5, 1.}

\begin{document}

\title{Where Galaxies Point: \\ {First Measurement of the Large-Scale Axial Intrinsic Alignment}}

\author{Pedro da Silveira Ferreira}
\email{dasferreira.pedro@gmail.com}
\affiliation{Center for Cosmology and Computational Astrophysics, Institute for Advanced Study in Physics, Zhejiang University, Hangzhou 310058, China.}
\affiliation{Institute of Astronomy, School of Physics, Zhejiang University, Hangzhou 310058, China}
\affiliation{Centro Brasileiro de Pesquisas F\'isicas, 22290-180, Rio de Janeiro, RJ, Brazil}
\affiliation{Observat\'orio do Valongo, Universidade Federal do Rio de Janeiro, 20080-090, Rio de Janeiro, RJ, Brazil}
\author{Rafael Oliveira Ramos}
\affiliation{Instituto de F\'isica, Universidade Federal do Rio de Janeiro, 21941-972, Rio de Janeiro, RJ, Brazil}
\author{Paula S. Ferreira}
\affiliation{Center for Cosmology and Computational Astrophysics, Institute for Advanced Study in Physics, Zhejiang University, Hangzhou 310058, China.}
\affiliation{Institute of Astronomy, School of Physics, Zhejiang University, Hangzhou 310058, China}
\affiliation{Observatório Nacional, 20921-400, Rio de Janeiro, RJ, Brazil}
\author{Arianna Cortesi}
\affiliation{Instituto de F\'isica, Universidade Federal do Rio de Janeiro, 21941-972, Rio de Janeiro, RJ, Brazil}
\affiliation{Observat\'orio do Valongo, Universidade Federal do Rio de Janeiro, 20080-090, Rio de Janeiro, RJ, Brazil}
\author{Fabricio Ferrari}
\affiliation{Instituto de Matemática Estatística e Física, Universidade Federal do Rio Grande, 96203-900, Rio Grande, RS, Brazil}
\author{Valerio Marra}
\affiliation{Departamento de Física, Universidade Federal de Ouro Preto, 35400-000, Ouro Preto, MG, Brazil}
\affiliation{INAF -- Osservatorio Astronomico di Trieste, via Tiepolo 11, 34131, Trieste, Italy}
\affiliation{IFPU -- Institute for Fundamental Physics of the Universe, via Beirut 2, 34151, Trieste, Italy\vspace{0.200cm}}
\author{Clécio R. Bom}
\affiliation{Centro Brasileiro de Pesquisas F\'isicas, 22290-180, Rio de Janeiro, RJ, Brazil}
\date{\today}
\author{Renyue Cen}
\email{renyuecen@zju.edu.cn}
\affiliation{Center for Cosmology and Computational Astrophysics, Institute for Advanced Study in Physics, Zhejiang University, Hangzhou 310058, China.}
\affiliation{Institute of Astronomy, School of Physics, Zhejiang University, Hangzhou 310058, China}

\begin{abstract}
We report evidence for large-scale axial intrinsic alignment (LAIA): a coherent axis shared by galaxies and cosmic-web filaments. Applying an orientation-field estimator to Dark Energy Survey (DES) Y3 shape data, we identify a preferred axis in galaxy orientations. Ellipticals' semi-major and spirals' semi-minor axes align with it, producing a $4.7\sigma$ signal whose pattern and amplitude hierarchy are consistent with morphology-dependent tidal-alignment and tidal-torquing expectations. Independently, Sloan Digital Sky Survey (SDSS) filament catalogues yield a compatible axis: northern and southern Galactic samples agree within $\simeq1\sigma$, the combined signal reaches $12.6\sigma$, and the axis lies within $\simeq2\sigma$ of the high-redshift galaxy sample direction. Because DES and SDSS footprints overlap marginally, this agreement is unlikely to arise from direct galaxy--filament alignment. It therefore provides a multi-survey, multi-observable test of a large-scale orientation field, stable under redshift and systematics tests. $N$-body mocks based on an isotropic $\Lambda$CDM cosmology with standard intrinsic-alignment prescriptions, including Euclid Flagship 2 and MICECAT v2, do not reproduce the pattern. LAIA provides a new statistical-isotropy probe linking galaxy morphology, cosmic-web structure and large-scale tidal fields.
\end{abstract}

\maketitle

\label{sec:introduction}

\noindent
Galaxies exhibit correlations in their orientations and shapes within their neighbourhood, known as intrinsic alignments (IA)~\cite{Chisari:2025gsy,Lamman:2023hsj}. These arise from tidal interactions with the surrounding large-scale structure (LSS), influencing galaxy formation and evolution. More broadly, the same gravitational field also shapes haloes, clusters and filaments, inducing alignments within the surrounding cosmic web. While weak lensing seeks to measure the distortions of galaxy shapes caused by intervening mass, IA can mimic or obscure lensing-induced correlations. This contamination is critical in cosmic shear studies used to infer cosmological parameters and the nature of dark energy~\cite{Hirata:2004gc,Joachimi:2015mma,Troxel:2014dba}.

Beyond cosmic shear, IA provides a window into galaxy formation and the large-scale tidal fields shaping galactic structure \cite{Crittenden:2000wi,Troxel:2014dba,Blazek:2015lfa,Chisari:2015qga,Codis:2015tla,Joachimi:2015mma,Kiessling:2015sma,Kirk:2015nma,Kraljic:2019lca,Samuroff:2020gpm}. Observations have established IA from arcminute-scale ellipticity correlations \citep{brown2002} to cluster-pair and galaxy-multiplet measurements over \(1\text{--}100\,\mathrm{Mpc}/h\) scales \citep{Smargon2012,desiIA}. These correlations constrain models of structure formation by encoding the interplay between dark matter haloes, baryonic processes, and the large-scale environment \cite{Kirk:2015nma,2012JCAP...05..041B}, and can probe primordial non-Gaussianity (PNG)~\cite{SchmidtChisariDvorkin2015,Kogai_2018}. Thus, while IA is often viewed as a contaminant in weak-lensing surveys, it also constitutes a valuable probe of the processes shaping galaxy evolution and the early Universe.

Looking ahead, surveys such as the China Space Station Telescope (CSST) \cite{zhan2021wide}, the Legacy Survey of Space and Time (LSST) \cite{collaboration2012large} and \textit{Euclid} \cite{scaramella2022euclid} will push weak-lensing measurements to higher precision, making IA modelling increasingly important. Interpreting these data will require a robust theoretical and simulation-based treatment of IA \cite{Joachimi:2015mma,Hilbert:2016ylf}. Useful starting points for such modelling are the nonlinear alignment (NLA) model \citep{Hirata:2004gc,BridleKing2007} and the tidal-alignment and tidal-torquing (TATT) framework \citep{Blazek2019}, which together describe IA responses and their scale and redshift dependence. The challenge is that IA is not a single physical effect, but a family of mechanisms whose relative importance depends on morphology, colour, mass and environment~\citep{Chisari:2025gsy,Lamman:2023hsj,Hirata:2004gc,Joachimi:2015mma,Troxel:2014dba,2011JCAP...05..010B,2012JCAP...05..041B,CatelanKamionkowskiBlandford2001,Peters:2024afu,Ghosh:2020zfa}.

\begin{figure*}[t]
    \includegraphics[width=0.71\linewidth]{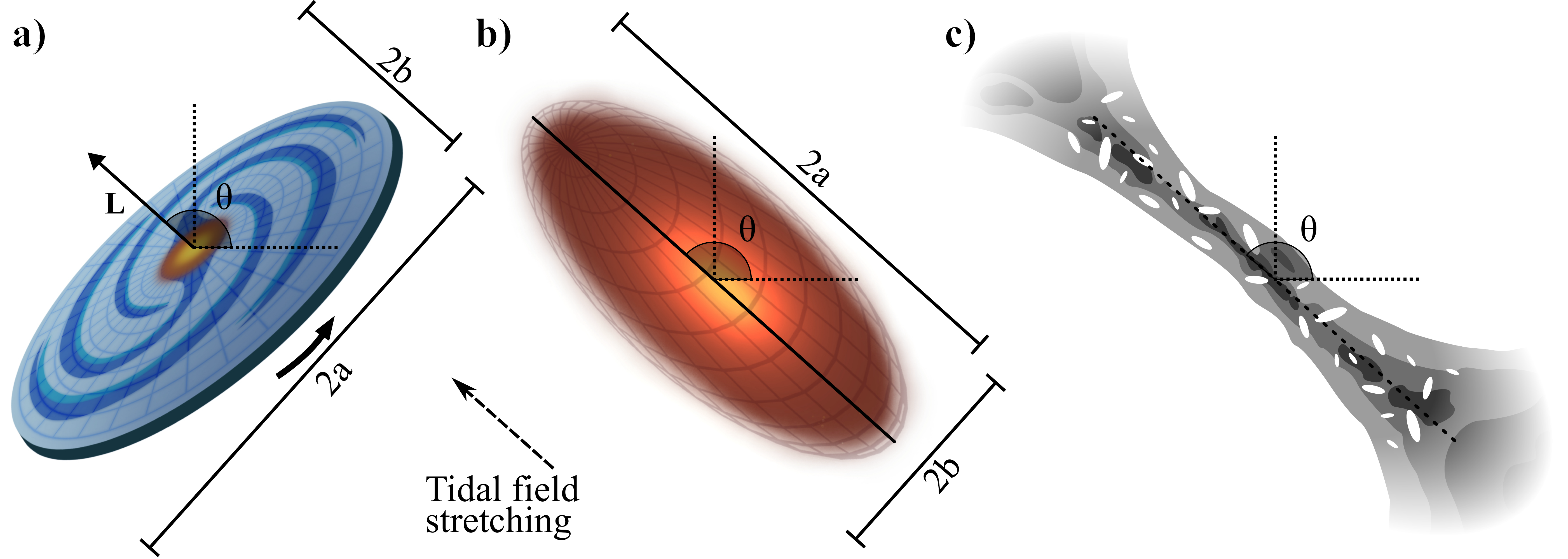}
    \caption{\textbf{Schematic illustration of the intrinsic alignment effect for different tracers.} \textbf{a)} Spiral galaxies tend to align their angular momentum vector $\mathbf{L}$ with the tidal field. The projection of $\mathbf{L}$ onto the image plane corresponds to the observed minor axis direction (2b direction) in the thin-disc limit. \textbf{b)} By contrast, elliptical galaxies typically align their major axes (2a direction). Thus, under the same tidal field, spirals and ellipticals tend to become orthogonal on average. \textbf{c)} Filament spines preferentially follow the stretching direction of the tidal field. LAIA searches for a common sky-scale axis among these orientation tracers, rather than for local galaxy-host filament alignment.}   
    \label{fig:ia_illustration}
\end{figure*}

We can go one step further and ask whether IA persists on even larger scales, up to the Universe as a whole. On such scales, a coherent anisotropic component of the gravitational potential could define a preferred direction in the LSS. The orientations of galaxies and cosmic-web filaments could respond to this common frame, producing an axial alignment. We refer to this effect as large-scale axial intrinsic alignment (LAIA). Unlike standard IA measurements based mainly on two-point correlations, LAIA treats orientations as a sky field and asks whether it contains a coherent axial component.

A conservative physical picture follows from the tidal-field physics underlying linear alignment and tidal torquing. In the linear-alignment picture, developed by \citet{CatelanKamionkowskiBlandford2001} and generalized in modern IA response formalisms \citep{Blazek2019}, pressure-supported galaxies trace the tidal tensor at the epoch of formation, whereas rotationally supported galaxies acquire angular momentum through tidal torquing. This makes the elliptical--spiral distinction both physical and observational: ellipticals are expected to respond mainly through a linear coupling that aligns their projected major axes with the tidal eigenframe, with the strongest signals in red, luminous and massive systems \citep{Troxel:2014dba,Joachimi:2015mma,Fortuna2021,Georgiou:2025pca}; spirals instead trace tidal torquing through spin correlations, for which the projected minor axis is a spin proxy in the thin-disc limit \citep{PadillaStrauss2008}. Because spin alignments arise from higher-order tidal couplings, their large-scale signal is generally expected to be weaker than the linear response of ellipticals \citep{CatelanKamionkowskiBlandford2001,Joachimi:2015mma,Blazek2019}. Filament spines are also shaped by tidal fields and tend to trace the stretching direction. Fig.~\ref{fig:ia_illustration} illustrates these tracer-dependent orientation responses. LAIA would in this sense provide a novel orientation-based probe of the tidal eigenframe. This picture also links LAIA to bulk-flow measurements, motivating a low-redshift comparison of their preferred directions. Although the two observables are sourced by different derivatives of the same large-scale gravitational potential, their axes can become approximately collinear when the local field is dominated by a large-scale attractor--repeller configuration, as suggested by local-flow reconstructions and galaxy--shear alignments \citep{Hoffman:2012,Tully:2014Nature,Hoffman:2017NatAs,Pahwa:2015}.

Galaxy orientations evolve through mergers, accretion, feedback and environment, which can weaken or erase halo-inherited alignments. Nevertheless, a residual large-scale memory may survive in galaxy morphologies and spin proxies. Filaments, by contrast, are a comparatively slowly evolving tracer and may retain a cleaner memory of the long-wavelength tidal eigenframe~\citep{BondKofmanPogosyan1996,Cautun:2014,GalarragaEspinosa:2024}. Their spines provide a third orientation tracer, with different astrophysical evolution and survey systematics. A common axis inferred from galaxies and filaments could then reflect two distinct tracers responding to the same large-scale orientation field, rather than a local galaxy--filament alignment.

A LAIA signal would therefore probe the cosmological principle, which states that the Universe is statistically homogeneous and isotropic on scales larger than about \(100\text{--}200\,\mathrm{Mpc}/h\) \citep{Abdalla:2022yfr,CosmoVerseNetwork:2025alb}. Preferred directions have been discussed in tests of the cosmological principle, including CMB anomalies, number count dipoles and dipole-like anisotropies in the SN Ia distance--redshift relation or in inferred~$H_0$ \citep{Perivolaropoulos:2021jda,Aluri:2022}. SN Ia Hubble-diagram and distance-ladder analyses have reported directional signals, with some rejecting isotropy at up to $\sim3$--$5\sigma$, although interpretation depends on redshift cuts, reference frame, peculiar-velocity corrections and sky coverage \citep{Colin:2019opb,Sorrenti:2022zat,Hu:2023eyf,Hu:2024qnx,Sah:2024csa}. Number count dipoles in radio galaxies and quasars have reached $\sim\!5\sigma$ tension with $\Lambda\rm{CDM}$ expectations \citep{Dalang:2021,Secrest:2022}. However, dipole observables remain mutually inconsistent, and some tensions may reflect tracer populations or selection effects \citep{FerreiraMarra2024,Abghari:2024}. At lower redshift, peculiar-velocity measurements are similarly unsettled: some report large coherent flows challenging $\Lambda$CDM, whereas others find standard-model amplitudes \cite{Watkins:2008,Watkins:2023,Scrimgeour:2015,Qin:2018,Tsagas:2025}. In this context, searching for LAIA offers a different perspective on these questions: a novel orientation-based test of statistical isotropy. In a related but distinct context, clustering--IA correlations have been forecast to probe primordial dipolar power modulation~\cite{Minato2025}.

These connections do not imply a direct amplitude mapping. While tidal-alignment models successfully describe large-scale IA, there is still no consolidated theoretical prediction for IA amplitudes on very-low-multipole angular scales~\cite{Kiessling:2015sma,Troxel:2014dba,2011JCAP...05..010B}. Accordingly, the LAIA signal cannot yet be translated into a unique amplitude for the underlying tidal-field anisotropy without an additional response model. We therefore frame the analysis phenomenologically: define an orientation statistic and calibrate its null distribution with realistic mocks.

\begin{figure*}[t] 
\includegraphics[width=0.79\textwidth]{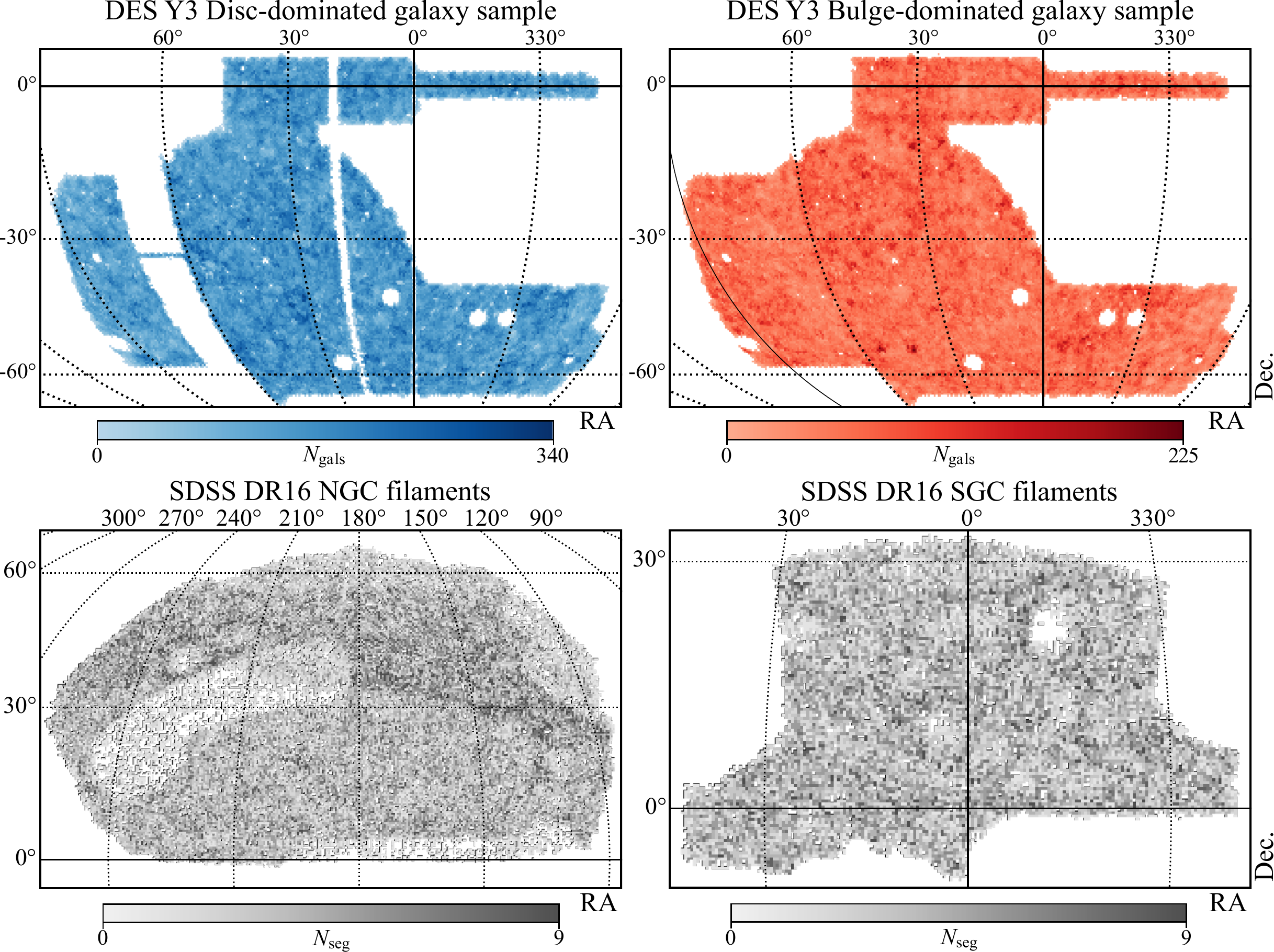}
\caption{\textbf{Survey footprints and tracer densities.} Mollweide projections in equatorial coordinates of the tracer-density maps used in this work, computed on a \texttt{HEALPix} grid with \texttt{NSIDE}=128. \textbf{Top left:} DES Y3 disk-dominated (DD; late-type/spiral) galaxies. \textbf{Top right:} DES Y3 bulge-dominated (BD; early-type/elliptical) galaxies. Differences between the DES galaxy footprints reflect the sky coverage of the corresponding morphology-classification catalogues \cite{DES:2020tkt,2021MNRAS.507.4425C}. \textbf{Bottom left:} SDSS DR16 Northern Galactic Cap (NGC) filament segments. \textbf{Bottom right:} SDSS DR16 Southern Galactic Cap (SGC) filament segments. Filament maps are built from skeleton segments of length $10\,h^{-1}\mathrm{Mpc}$.}
\label{fig:ds_bs_density_map}\vspace{0.27cm}
\end{figure*}

More speculatively, a large-scale axial pattern could reflect primordial physics that breaks statistical isotropy, including anisotropic PNG \citep{SchmidtChisariDvorkin2015}. Detecting LAIA would therefore be complementary to CMB- and LSS-based probes and sensitive to models with direction-dependent initial conditions. Here we introduce a LAIA estimator and apply it to Dark Energy Survey (DES) Year 3 galaxy shapes and Sloan Digital Sky Survey (SDSS) filament catalogues, providing the first measurement of this signal.

\section*{\label{sec:cosmicaxis} Searching for a cosmic axis}

\begin{figure}
\includegraphics[width=0.93\linewidth]{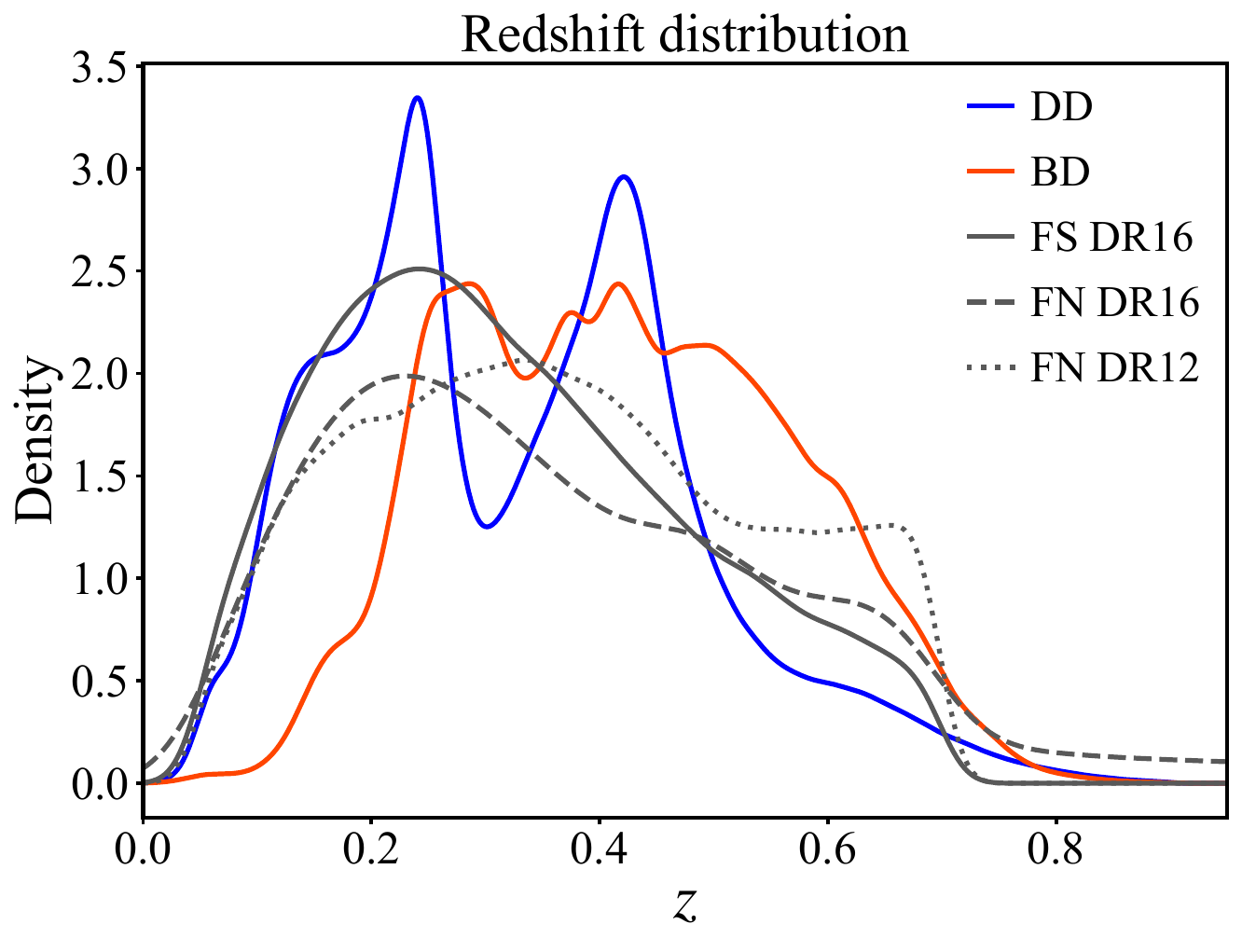}
\caption{\textbf{Redshift distributions.} Photometric redshift distributions of the DES Y3 disk-dominated (DD) and bulge-dominated (BD) galaxy samples, and spectroscopic redshift distributions of the centres of the SDSS DR16 NGC (FN DR16), SDSS DR16 SGC (FS DR16) and SDSS DR12 NGC (FN DR12) filament segments.}\label{fig:z_distribution}
\end{figure}

To search for LAIA, we translate the idea of a coherent orientation field into a measurement framework, specifying the tracers, estimator and null tests used below. In practice, we divide galaxies into bulge-dominated (BD) and disk-dominated (DD) systems, since pressure support versus rotational support sets the operative mechanism. To obtain classified samples with precise position angles (PAs), we use the DES Y3 shape catalogue \cite{DES:2020ekd}, whose ellipticity components are estimated with the \texttt{metacalibration} algorithm \cite{DES:2020ekd}. This catalogue is cross-matched with the morphological classifications made by \citet{DES:2020tkt} and \citet{2021MNRAS.507.4425C}, and bulge-to-total fraction constraints \cite{DES:2020aks}. The resulting analysis samples contain 3\,007\,638 DD and 1\,381\,049 BD galaxies. For filaments, we use the SDSS DR16 filament catalogue of \citet{Duque:2021xgw} as our main sample, separated into Northern Galactic Cap (NGC) subset (FN DR16) and Southern Galactic Cap (SGC) subset (FS DR16), and the SDSS DR12 filament catalogue of \citet{Chen:2015bra}, FN DR12 sample, which covers only the NGC, as a consistency check. We represent the reconstructed filament spines by segments of fixed comoving length $10\,h^{-1}{\rm Mpc}$ and use their PAs as the filament-orientation tracers. The final filament-segment samples comprise 104\,517 FN DR16, 36\,623 FS DR16 and 97\,598 FN DR12 segments. Unlike the galaxy samples, these filament orientations are not image-level shape measurements and therefore do not require PSF-, resolution- or ellipticity-based quality cuts; residual footprint and sampling effects are instead assessed with the filament null catalogues. The complete sample selection and segment construction are described in Methods~\ref{app:data}. See Fig.~\ref{fig:ds_bs_density_map} for the \texttt{HEALPix}~\cite{Gorski:2004by} density maps and Fig.~\ref{fig:z_distribution} for the redshift distributions.

Operationally, LAIA is defined as a sky-scale axial component of an orientation field, characterized by a preferred axis and an amplitude. For each orientation tracer $X$, we search for the sky axis $\hat{\mathbf d}$ that maximizes the coherent alignment of headless position angles. Here, $X=BD$ denotes the bulge-dominated galaxies, $X=DD$ the disk-dominated galaxies, and $X=F$ filament segments. Denoting the corresponding headless orientation by $\hat{\mathbf u}^{X}_i$, with no intrinsic arrow ($\hat{\mathbf u}^{X}_i\equiv-\hat{\mathbf u}^{X}_i$), and its sky position by $\hat{\mathbf n}^{X}_i$, with $i$ indexing galaxies or filament segments, we use the squared projection of $\hat{\mathbf u}^{X}_i$ onto the local-sky-plane projection of $\hat{\mathbf d}$ as a quadratic measure of axial alignment. Thus, the LAIA of tracer $X$ can be measured with
\begin{equation}\label{eq:estimator}
\hat{E}^{X}(\hat{\mathbf d})
=
\sum_{i=1}^{N_X}
\frac{
w^{X}_i\bigl(\hat{\mathbf d}\cdot\hat{\mathbf u}^{X}_i\bigr)^2
}{
1-\bigl(\hat{\mathbf d}\cdot\hat{\mathbf n}^{X}_i\bigr)^2
},
\qquad X\in\{BD,DD,F\},
\end{equation}
where the denominator normalizes by the squared norm of the projection of $\hat{\mathbf d}$ onto the local tangent plane. For galaxies, $\hat{\mathbf u}^{BD}_i=\hat{\mathbf a}_i$ (semi-major axes) and $\hat{\mathbf u}^{DD}_i=\hat{\mathbf b}_i$ (semi-minor axes); for filaments, $\hat{\mathbf u}^{F}_i=\hat{\mathbf f}_i$ is the projected filament-spine direction. We assign normalized inverse-variance PA weights to galaxies,
\begin{equation}\label{w_error}
    w_i^X=\frac{(\sigma_{\theta,i}^{X})^{-2}}{\displaystyle\sum_{j=1}^{N_X}(\sigma_{\theta,j}^{X})^{-2}},
    \qquad X\in\{BD,DD\},
\end{equation}
where $\sigma_{\theta,i}^{X}$ denotes the PA uncertainty (Methods~\ref{app:theta_error}). For filaments, no PA weighting is applied; we use equal weights $w_i^F=1/N_F$. Possible reconstruction-dependent effects are assessed by comparing the SDSS DR16 and DR12 filament catalogues, which use related but distinct methods. We evaluate $\hat{E}^{X}(\hat{\mathbf d})$ over trial axes using a gradient-ascent optimization algorithm. The axis that maximizes $\hat{E}^{X}(\hat{\mathbf d})$ defines the LAIA direction $\hat{\mathbf d}_{\rm IA}^{X}$ of the tracer; see Methods~\ref{app:laia} for the complete estimator definition.

To express the alignment strength as an angular amplitude, consider a simple phenomenological model in which a long-wavelength tidal field present during structure formation imprints a nearly uniform deviation from random position angles across the sample. Rather than describing a dynamical rotation of individual objects, the model parameterizes the observed orientation distribution as a coherent displacement of headless axes toward the recovered LAIA direction, $\hat{\mathbf d}_{\rm IA}^{X}$. For BD and DD galaxies, this corresponds to an average displacement of the semi-major and semi-minor axes, respectively; for filaments, it corresponds to an effective alignment angle of the projected spine directions. Under this common orientation model, the LAIA amplitude $\theta_{\rm IA}^{X}$ can be written (see Methods~\ref{app:estimator}) as
\begin{equation}\label{eq:theta_ia_sol}
2\theta_{\rm IA}^{X}+\sin\bigl(2\theta_{\rm IA}^{X}\bigr)=
\frac{2\pi}{\overline{C}_{X}}
\Bigl(\hat{E}^{X}(\hat{\mathbf d}_{\rm IA}^{X})-\frac{1}{2}\Bigr),
\end{equation}
where, for galaxy position angles,
\begin{equation}
\overline{C}_{X}=\sum_{i=1}^{N_X}w^{X}_i\, e^{-2(\sigma_{\theta,i}^{X})^2},
\qquad X\in\{BD,DD\}.
\end{equation}
For filaments, we set $\overline{C}_{f}=1$. Here $\theta_{\rm IA}^{X}$ quantifies how strongly the position angles of tracer $X$ depart from the null orientation distribution toward the axis.

Having defined the estimator, we calibrate its null behaviour separately for galaxy shapes and filament orientations. For galaxies, we compare the observed orientation field with $N$-body-based mock catalogues with IA prescriptions, including Euclid Flagship 2 \cite{Euclid:2024few,Euclid:2026bur,Potter:2016ttn,2016ascl.soft09016P,Behroozi:2011ju} (Euclid FS2) and MICECAT v2 \cite{DES:2022aeh,mice1,mice2,mice3,mice4,mice5} (Methods~\ref{app:nbody_sampling}; see Extended Data Fig.~\ref{fig:nbody_sampling} for details of the sampled distributions), and with DES-based random realizations generated from residual ellipticity components after modelling and subtracting correlations with PSF and survey-condition maps (Methods~\ref{sec:residual_correlations}; see Extended Data Figs.~\ref{fig:bulge_e_sys_raw}-\ref{fig:disc_e_sys_residual} for observed systematic correlations and residuals after corrections). For filaments, we use binned-permutation nulls, grouping parent filaments by local sky patch ($\sim15^\circ \times 15^\circ$) and redshift bin. Within each bin, their axial orientations are permuted, preserving the footprint, redshift distribution, local orientation distribution and internal segment coherence, while breaking the large-scale association between sky position and filament orientation (see Methods~\ref{app:filament_nulls}). These tests address complementary null hypotheses: for galaxies, whether standard IA prescriptions in current $\Lambda$CDM $N$-body mock catalogues or residual survey systematics can reproduce the orientation field; for filaments, whether the segment-orientation field contains sky-scale coherence beyond the footprint, selection function and local sky/redshift-dependent orientation statistics. As an additional consistency check, LAIA-injected simulations with varying angular coherence scales show that the estimator responds weakly unless the alignment remains coherent over scales of order $\gtrsim30^\circ$ (Methods~\ref{sec:coherence}; see Extended Data Fig.~\ref{fig:coherence}). 

In the tidal-field picture motivating LAIA, compatibility between the axes recovered from BD galaxies, DD galaxies and filament spines would then provide a multi-tracer consistency test of a common large-scale orientation field. 

\begin{figure*}[t]
    \includegraphics[width=0.821\textwidth]{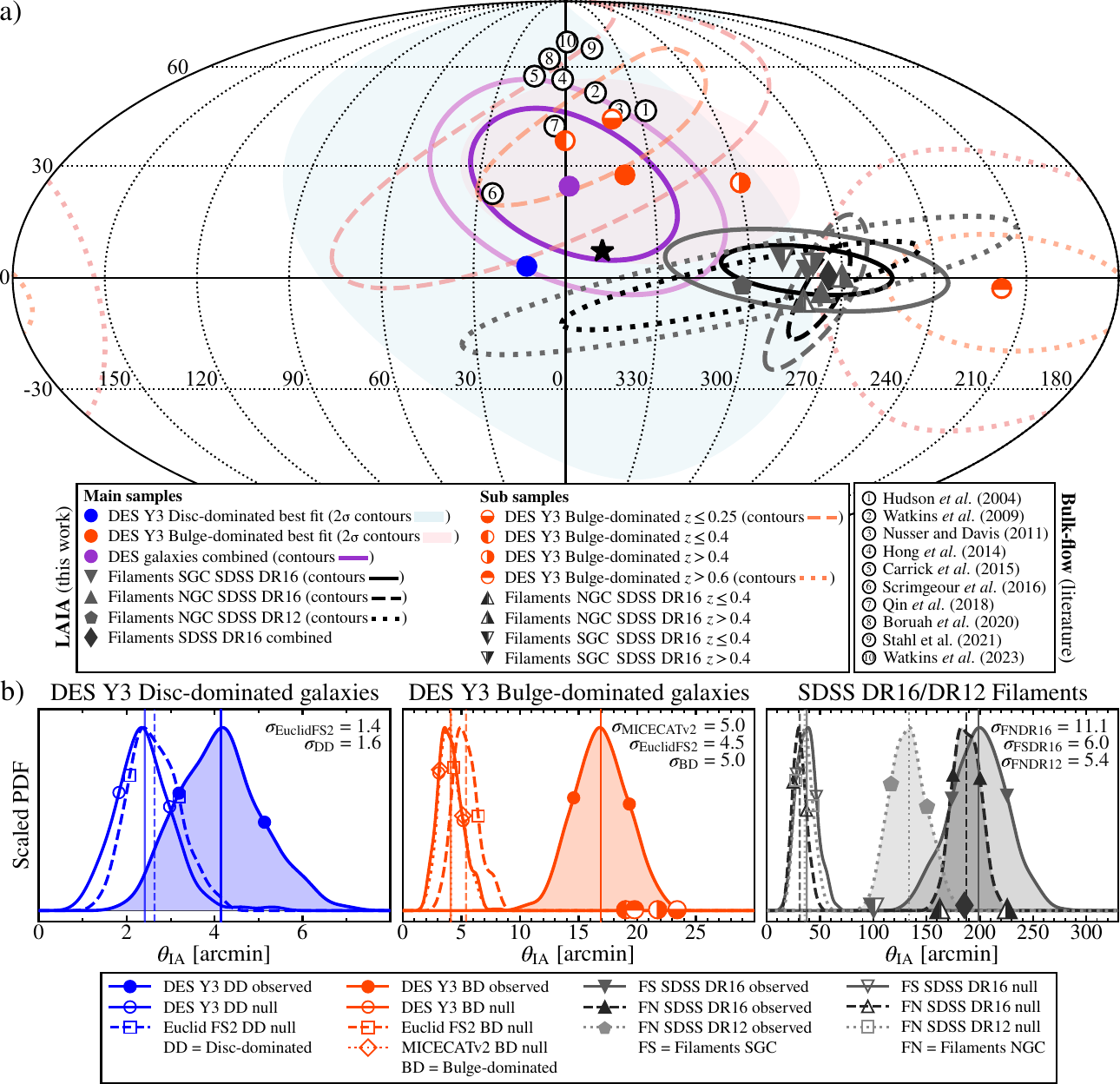}
\caption{\textbf{LAIA directions and amplitudes.}
\textbf{a)} Highest-density interval contours and best-fit LAIA axes, $\hat{\mathbf d}_{\rm IA}$, in equatorial coordinates for DES Y3 bulge-dominated galaxies (BD; red), DES Y3 disk-dominated galaxies (DD; blue), the combined BD+DD directional likelihood (purple) and SDSS filament samples (grey). Because the estimator is axial, $\hat{\mathbf d}_{\rm IA}\equiv-\hat{\mathbf d}_{\rm IA}$; we therefore show one representative direction for each axis, with the equivalent antipodal direction given by $(\alpha+180^\circ,-\delta)$. Markers denote best-fit directions. Low-redshift BD axes are directionally compatible with several published bulk-flow directions, shown as numbered circles in the antipodal hemisphere. Numbers 1--10 correspond, in order, to Refs.~\citep{Hudson:2004et,Watkins:2008,Nusser:2011tu,Hong:2014jla,Carrick:2015xza,Scrimgeour:2015,Qin:2018,Boruah:2019icj,Stahl:2021mat,Watkins:2023}. When a study reports multiple estimates, we display the highest-redshift measurement. High-redshift BD measurements move toward the filament LAIA axis, while the filament axes remain stable across the tested redshift splits. The antipodal CMB dipole direction is shown with the $\bigstar$ symbol.
\textbf{b)} Effective LAIA amplitude, $\theta_{\rm IA}$, for the main tracer samples and their null distributions. Filled distributions show the observed $\theta_{\rm IA}$ distributions obtained from 500 spatial block-bootstrap resamples, whereas unfilled distributions show the corresponding null ensembles. Vertical lines (full samples) and markers near the $\theta_{\rm IA}$ axis (sub samples) indicate the measured $\theta_{\rm IA}$. Additional markers on the distributions indicate the quantiles corresponding to the $\pm1\sigma$ intervals. For galaxies, null distributions are obtained from Euclid Flagship 2 and MICECAT v2.0 $N$-body-based catalogues built in statistically isotropic $\Lambda$CDM cosmologies with standard IA prescriptions; these mocks are resampled 500 times, separately by morphology, to match the DES joint photometric-redshift--$i$-band-magnitude distribution and PA-error distribution, yielding the $\sigma_{\rm EuclidFS2}$ and $\sigma_{\rm MICECATv2}$ significances. DES-based empirical nulls, built from residual ellipticity components after removal of PSF and survey-condition correlations, yield the $\sigma_{\rm BD}$ and $\sigma_{\rm DD}$ significances. MICECAT v2.0 is shown for BD only as its IA implementation does not include spiral-galaxy IA. For filaments, null distributions are obtained from binned permutations of the filament orientations. When multiple nulls are available, the significance quoted in the text is the most conservative.}
    \label{fig:results}
\end{figure*}

\section*{\label{sec:discussion} Results}

\providecommand{\nodata}{--}
\providecommand{\rotsect}[2][0]{\smash{\raisebox{\dimexpr-0.50\height-#1\baselineskip\relax}{\rotatebox{90}{\textbf{#2}}}}}
\begin{table}[t]
\centering
\label{tab:laia_summary}
\scriptsize
\setlength{\tabcolsep}{1.1pt}
\renewcommand{\arraystretch}{1.35}
\resizebox{\columnwidth}{!}{%
\begin{tabular}{@{}c@{\hspace{0.25em}}l@{\hspace{0.35em}}c@{\hspace{0.35em}}c@{\hspace{0.35em}}c@{\hspace{0.55em}}c@{\hspace{0.35em}}c@{\hspace{0.35em}}c@{}}
\toprule\toprule
 & & \multicolumn{3}{c}{\textbf{Bulge-dominated}} & \multicolumn{3}{c}{\textbf{$\;\;\;\;\;\;$Disk-dominated$\;\;\;\;\;\;$}} \\
\cmidrule(r{0.30em}){3-5}\cmidrule(l{0.30em}){6-8}
 & \textbf{Sample} & $\;\;\theta_{\rm IA}$ ($'$) \;& $\alpha$ ($^\circ$) & $\delta$ ($^\circ$) & $\;\;\;\theta_{\rm IA}$ ($'$)\; & $\alpha$ ($^\circ$) & $\delta$ ($^\circ$) \\
\midrule
 & $z\leq0.25$ & $23^{+8}_{-1}$ & $342^{+32}_{-33}$ & $44^{+15}_{-16}$ & \nodata & \nodata & \nodata \\
 & $z\leq0.4$ & $19^{+4}_{-3}$ & $0^{+28}_{-28}$ & $37^{+15}_{-14}$ & \nodata & \nodata & \nodata \\
\rotsect[0.50]{Galaxies DES $\;\;\;$} & $z>0.4$ & $22^{+3}_{-3}$ & $300^{+37}_{-35}$ & $25^{+16}_{-34}$ & \nodata & \nodata & \nodata \\
 & $z>0.6$ & $20^{+4}_{-4}$ & $218^{+23}_{-30}$ & $-3^{+9}_{-11}$ & \nodata & \nodata & \nodata \\
 & Full & $17^{+2}_{-2}$ & $339^{+22}_{-24}$ & $28^{+13}_{-8}$ & $4^{+1}_{-1}$ & $13^{+45}_{-44}$ & $3^{+58}_{-17}$ \\
\cmidrule{2-8}
 & Combined & \multicolumn{6}{c}{$(\alpha,\delta)=(359^{+23}_{-24},\,24^{+15}_{-13})^\circ$} \\
\midrule\midrule
 & & \multicolumn{3}{c}{\textbf{NGC DR16}} & \multicolumn{3}{c}{\textbf{SGC DR16}} \\
\cmidrule(r{0.30em}){3-5}\cmidrule(l{0.30em}){6-8}
 & \textbf{Sample} & $\;\;\theta_{\rm IA}$ ($'$)\;\; & $\alpha$ ($^\circ$) & $\delta$ ($^\circ$) & $\;\;\;\theta_{\rm IA}$ ($'$)\; & $\alpha$ ($^\circ$) & $\delta$ ($^\circ$) \\
\midrule
 & $z\leq0.4$ & $163^{+13}_{-13}$ & $282^{+12}_{-12}$ & $-6^{+19}_{-19}$ & $100^{+21}_{-18}$ & $289^{+63}_{-59}$ & $5^{+28}_{-32}$ \\
 & $z>0.4$ & $226^{+16}_{-18}$ & $271^{+11}_{-11}$ & $0^{+8}_{-8}$ & $420^{+39}_{-31}$ & $279^{+14}_{-14}$ & $4^{+3}_{-3}$ \\
\rotsect{Filaments SDSS $\;\;\;\;\;\;\;$} & Full & $187^{+13}_{-12}$ & $277^{+6}_{-6}$ & $-4^{+8}_{-8}$ & $199^{+26}_{-26}$ & $281^{+19}_{-18}$ & $2^{+4}_{-4}$ \\
\cmidrule{2-8}
 & Combined & \multicolumn{6}{c}{$\theta_{\rm IA}=187^{+10}_{-9}\,{'};\;(\alpha,\delta)=(274^{+5}_{-4},\,0^{+3}_{-4})^\circ$} \\
\cmidrule{2-8}
 & & \multicolumn{3}{c}{\textbf{NGC DR12}} & \multicolumn{3}{c}{} \\
\cmidrule{3-5}
 & Full & $133^{+18}_{-16}$ & $302^{+39}_{-38}$ & $-2^{+8}_{-8}$ & \nodata & \nodata & \nodata \\
\bottomrule\bottomrule
\end{tabular}%
}
\vspace{2pt}
\begin{minipage}{\columnwidth}
\caption{\label{tab:results}
\textbf{Summary of LAIA amplitudes and axes.}
Effective LAIA amplitudes, $\theta_{\rm IA}$, and preferred axial directions, $\hat{\mathbf d}_{\rm IA}$, for the full samples and subsamples. Uncertainties are marginal $68.27\%$ highest-density intervals. Directions are reported in equatorial coordinates, with right ascension $\alpha$ and declination $\delta$ in degrees; $\theta_{\rm IA}$ is reported in arcminutes. Because LAIA is axial, $\hat{\mathbf d}_{\rm IA}\equiv-\hat{\mathbf d}_{\rm IA}$, the listed direction is chosen in the same hemisphere as Fig.~\ref{fig:results}; the antipode $(\alpha+180^\circ,-\delta)$ is equivalent. Bulge-dominated galaxy measurements use semi-major axes, disc-dominated galaxy measurements use semi-minor axes, and filament measurements use projected spine directions. For the combined bulge--disc result, the likelihoods are combined only in direction because the two morphologies have different response amplitudes.
}
\end{minipage}
\end{table}

The LAIA measurements for the full galaxy and filament samples and their subsamples are summarized in Fig.~\ref{fig:results} and Table~\ref{tab:results}. The upper panel of Fig.~\ref{fig:results} shows the recovered axes in equatorial coordinates and confidence contours. Because LAIA is axial, $\hat{\mathbf d}_{\rm IA}\equiv-\hat{\mathbf d}_{\rm IA}$, we display only one representative direction for each measurement; the equivalent antipodal direction is $(\alpha+180^\circ,-\delta)$. Angles are reported in degrees for the sky coordinates $(\alpha,\delta)$ and in arcminutes for the effective alignment amplitude $\theta_{\rm IA}$. The lower panel compares measured amplitudes with null distributions: Euclid FS2 and MICECAT v2.0 mocks with standard IA prescriptions and DES-like PA errors for the galaxy samples, DES-based empirical nulls from residual ellipticity components after removal of PSF and survey-condition correlations, and binned-permutation nulls for the filament catalogues. For galaxies, the reported LAIA measurements are corrected for a residual RA-dependent declination bias (Methods~\ref{sec:residualbias}; see Extended Data Fig.~\ref{fig:residual_bias}) and for misclassification bias (Methods~\ref{sec:misclassification}); these corrections are calibrated with LAIA-injection tests in the DES-based empirical null catalogues (Methods~\ref{sec:residual_correlations} and \ref{sec:naive_mocks}). Contours and uncertainties for all tracers are obtained with a spatial block bootstrap (Methods~\ref{sec:error}). We also account for lensing leakage, calibrated using BUZZARD $N$-body simulations \cite{buzzard2,buzzard1} (Methods~\ref{sec:lensing}) and apply estimator-level PSF and baseline corrections (Methods~\ref{sec:psf}). See Methods~\ref{sec:pipeline} and Extended Data Fig.~\ref{fig:pipeline} for the pipeline summary.

For the DES Y3 galaxy samples, the bulge-dominated (BD) and disk-dominated (DD) catalogues show the morphology-dependent pattern expected from tidal-alignment and tidal-torquing physics. The full DD sample yields a weak but directionally compatible signal, $\theta_{\rm IA}=4\pm1{}'$. This is expected because spin-related alignments arise from higher-order tidal couplings and are generally weaker than linear alignment. By contrast, the full BD sample shows a much stronger signal, $\theta_{\rm IA}=17\pm2{}'$. Relative to the Euclid FS2 null, the corresponding significances are $1.4\sigma$ and $4.5\sigma$, respectively. The BD semi-major-axis direction and DD semi-minor-axis direction are compatible at $\sim1\sigma$, allowing a combined galaxy-axis estimate. Under the LAIA hypothesis, the preferred axis is a common parameter of the large-scale orientation field, whereas the response amplitude is tracer-dependent (tidal-alignment vs tidal-torquing). We therefore combine only the directional likelihoods, obtaining a shared preferred axis $(\alpha,\delta)=(359^{+23}_{-24},\,24^{+15}_{-13})^\circ$ with a combined significance of $4.7\sigma$.

The filament catalogues reveal a stronger and internally consistent axial pattern. The SDSS DR16 NGC and SGC filament samples yield significances of $11.1\sigma$ and $6.0\sigma$, respectively, and their recovered axes are mutually compatible in both direction and amplitude. Combining the two DR16 filament samples gives $\theta_{\rm IA}=187^{+10}_{-9}\,{}'$, $(\alpha,\delta)=(274^{+5}_{-4},\,0^{+3}_{-4})^\circ$, with a significance of $12.6\sigma$ relative to the binned-permutation filament null. The independent SDSS DR12 NGC filament catalogue provides a consistent cross-check, recovering a compatible preferred axis within the uncertainties. Because the DR16 NGC and SGC samples occupy disjoint sky regions, their agreement is not driven by shared environments.

We next test whether the recovered LAIA axes vary with redshift. For BD galaxies and for the DR16 NGC and SGC filament samples, we split the data at $z=0.4$. For BD galaxies, we also consider the more extreme low- and high-redshift cuts $z\leq0.25$ and $z>0.6$. We do not report redshift splits for the DD sample: its full-sample signal is already weak, leaving the corresponding subsamples statistically uninformative and the resulting measurements unstable. The BD axis shows a clear trend: as the high-redshift cut is raised, the recovered direction moves toward the filament axis, while the lowest-redshift BD measurement approaches the reported directions of large-scale bulk-flow estimates. In Fig.~\ref{fig:results}, the bulk-flow directions are shown in the antipodal hemisphere corresponding to our axial convention. Five of the literature bulk-flow directions are compatible with the $z\leq0.25$ BD axis within $1\sigma$, another two within $2\sigma$, and three marginally outside $2\sigma$. By contrast, the filament axes show no significant variation between $z\leq0.4$ and $z>0.4$, consistent with the interpretation of filaments as a more stable tracer of the large-scale tidal eigenframe.

Taken together, the measurements establish three empirical features: 
\vspace{-0.1cm}
\begin{enumerate}
    \item DES galaxy orientations contain a preferred axial component that is not reproduced by standard local IA prescriptions in current $N$-body-based catalogues built in statistically isotropic $\Lambda$CDM cosmologies;\vspace{-0.2cm}
    \item SDSS filament spines independently trace a sky-scale IA with high significance;\vspace{-0.2cm}
    \item The redshift dependence of the galaxy signal suggests a transition from local-flow-related directions at low redshift to a common large-scale tidal frame traced by both galaxies and filaments at higher redshift.
\end{enumerate}
\vspace{-0.1cm}
These results define the observational basis for LAIA. 

\section*{\label{sec:Conclusion} Discussion and conclusions}

We have presented the first evidence for a large-scale axial intrinsic alignment (LAIA): the semi-major axes of bulge-dominated (elliptical) galaxies and the semi-minor axes of disk-dominated (spiral) galaxies coherently point toward a common direction, with amplitudes obeying the hierarchy expected from tidal-alignment/torquing physics. In disk-dominated systems, the signal suggests possible long-term coherence in the projected angular-momentum direction. In bulge-dominated systems, different redshift ranges indicate evolution in galaxy orientations, from an axis compatible with the filament-traced direction at high redshift ($z>0.4$--$0.6$; $\simeq2\sigma$) to one approaching the reported cosmic bulk-flow direction at $z\leq0.25$ ($\simeq1\sigma$), which may reflect the growing influence of nearby large-scale structures. Although the lower-redshift subsample is not strictly local, directional statistics can be disproportionately affected by anisotropies in the surrounding matter distribution. If the local gravitational field is dominated by a large-scale attractor--repeller axis, as suggested by local-flow reconstructions \citep{Tully:2014Nature,Hoffman:2017NatAs}, a tidal-alignment axis correlated with the bulk-flow direction is physically plausible (Methods~\ref{app:bulkflow_laia_relation}). Filaments behave differently: consistent with spines tracing a comparatively slowly evolving cosmic-web skeleton \citep{BondKofmanPogosyan1996,Cautun:2014,GalarragaEspinosa:2024}, their LAIA direction shows negligible variation between $z\leq0.4$ and $z>0.4$. 
Remarkably, the Galactic northern and southern filament samples, drawn from non-overlapping light cones, recover compatible axes, as do independent filament catalogues. It is difficult to reconcile these results with survey artefacts. A conservative interpretation is that LAIA traces a long-wavelength tidal field acting on galaxies and filaments across horizon-scale modes. Together, these measurements provide a coherent phenomenological picture of the large-scale tidal orientation field. 

Compared to Euclid Flagship 2 and MICECAT v2.0 $N$-body galaxy mocks, the observed galaxy LAIA signal is inconsistent with standard local IA prescriptions at $4.7\sigma$. For filaments, the combined signal reaches $12.6\sigma$ relative to our empirical filament null based on binned permutations. However, a residual systematic capable of rotating both galaxy samples in phase cannot be completely excluded. Such an effect would mainly bias the recovered preferred direction, but would not naturally reproduce the morphology-dependent axial pattern, redshift behaviour, or the independent filament alignment.

This measurement has become possible only with modern wide-area imaging capable of resolving morphologies and position angles to the required accuracy for millions of galaxies, alongside spectroscopic surveys that enable precise filament reconstruction. In the near future, surveys such as CSST, \textit{Euclid}, LSST, DESI \cite{DESI:2013agm}, and PFS \cite{pfssurvey} will sharpen these measurements and allow tomographic LAIA analyses with finer redshift resolution.

A key next step is to extend simulations and models to predict axial alignments on horizon-scale modes and calibrate the LAIA amplitude response to the underlying tidal field, connecting the amplitude hierarchy of ellipticals and spirals to tidal-torquing expectations, galaxy assembly, cosmic-web geometry and possible early-Universe physics. Cross-correlations of LAIA with weak-lensing shear, CMB lensing, and forthcoming 21\,cm tracers could provide independent checks on systematics and help isolate the underlying tidal field. By turning galaxy and filament orientations into a precision probe on the largest scales, LAIA offers a bridge between galaxy evolution and cosmology and a new way to test statistical isotropy.

\vspace{1.2\baselineskip}
\noindent\rule{\columnwidth}{0.4pt}
\vspace{-1.5\baselineskip}

\setcounter{figure}{0}

\makeatletter
\renewcommand{\theHfigure}{ED.\arabic{figure}}

\renewcommand{\fnum@figure}{Extended Data Figure~\thefigure}%
\makeatother

\section*{Methods}
\label{sec:methods}

\subsection{\label{sec:pipeline}Galaxy-analysis pipeline}

\begin{figure}
\vspace{1.2mm}
\includegraphics[width=0.905\linewidth]{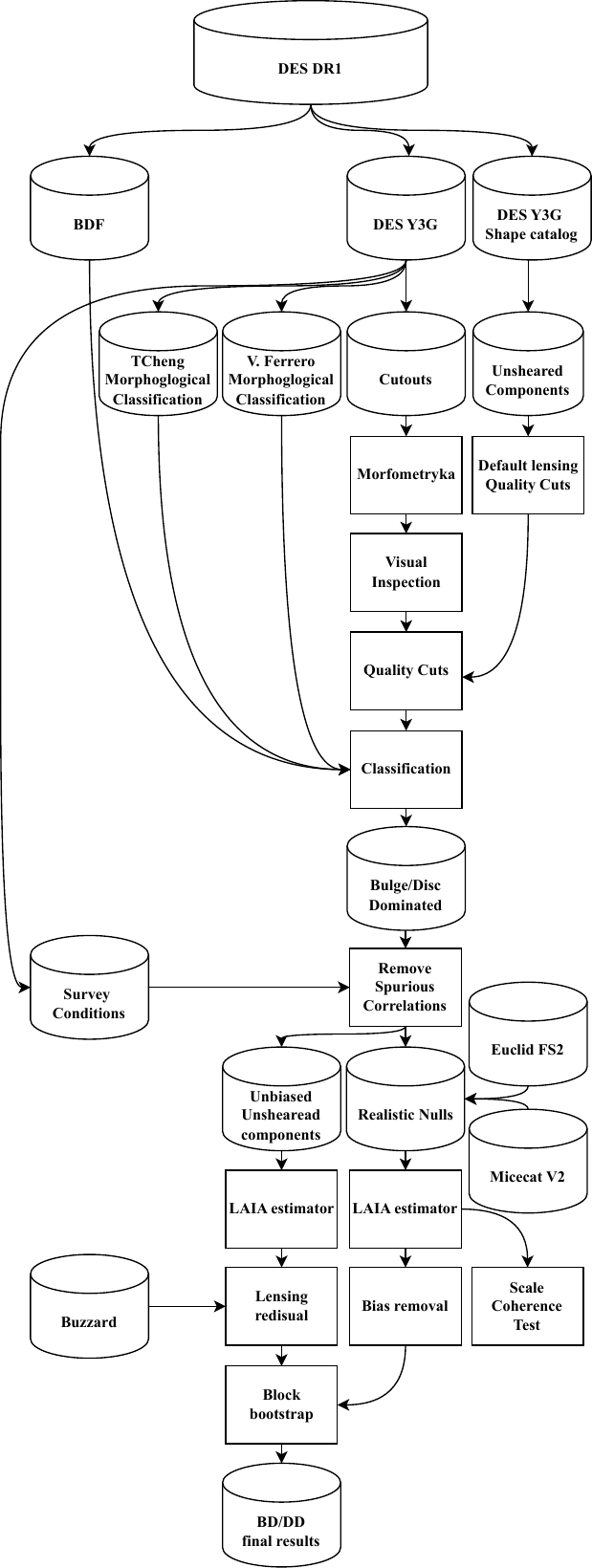}
\caption{
\textbf{Galaxy LAIA analysis pipeline.}
Schematic flowchart of the DES galaxy pipeline leading to the final bulge-dominated and disc-dominated LAIA measurements. Cylinders denote input catalogues, derived data products, mock catalogues or null catalogues, whereas rectangles denote selection, correction, validation or measurement steps.}
\label{fig:pipeline}
\end{figure}

The galaxy pipeline is the most involved part of the analysis because it combines image-level shape measurements, morphology and bulge-to-total information, quality cuts, residual-systematics removal, empirical null construction, mock-catalogue comparisons and injection-based validation tests. Extended Data Figure~\ref{fig:pipeline} summarizes the galaxy-analysis pipeline used to obtain the DES Y3 LAIA measurements.

The pipeline starts by constructing the bulge-dominated and disc-dominated samples from the DES Y3 shape catalogue, cross-matched with the DES DR1-based morphology and bulge-to-total products, after applying the selection and PA-quality cuts described in Methods~\ref{app:data}. We then remove residual correlations with PSF and survey-conditions and construct DES-based empirical null catalogues from the residual ellipticity components (Methods~\ref{sec:residual_correlations}). In parallel, Euclid Flagship 2 and MICECAT v2.0 mock catalogues provide external null tests based on standard local IA prescriptions in statistically isotropic $\Lambda$CDM cosmologies (Methods~\ref{app:nbody_sampling}).

The LAIA estimator is applied to the corrected data, empirical nulls and mock catalogues. Additional validation steps quantify lensing leakage (Methods~\ref{sec:lensing}), estimator-level PSF contamination and baseline correction (Methods~\ref{sec:psf}), coherence-scale response (Methods~\ref{sec:coherence}), residual directional bias (Methods~\ref{sec:residualbias}) and morphology-misclassification bias (Methods~\ref{sec:misclassification}). These injection-based tests (coherence-scale response, residual directional bias and morphology-misclassification bias) use the LAIA injection pipeline described in Methods~\ref{sec:naive_mocks}. Final confidence contours and uncertainties are obtained with the spatial block bootstrap described in Methods~\ref{sec:error}.

The filament analysis is substantially less involved and is therefore not shown as a separate flowchart. Filament orientations are measured from reconstructed, extended cosmic-web spines rather than from image-level galaxy shapes; because they are large-scale structural tracers, they do not require PSF-leakage corrections, PA-quality cuts, morphology classification, bulge-to-total selection, or subtraction of PSF- and survey-condition-correlated ellipticity components. Their relevant systematics are different: footprint and redshift selection, skeleton sampling, local orientation anisotropies and dependence on the filament reconstruction method. We address these effects by representing spines as fixed-length $10\,h^{-1}{\rm Mpc}$ segments, using binned-permutation nulls that preserve local sky and redshift structure, and repeating the measurement with independent SDSS DR16 and DR12 filament catalogues based on different reconstruction methods. The corresponding data selection, filament null construction, LAIA-injection validation, residual-bias checks and bootstrap uncertainty estimates are described in Methods~\ref{app:data}, \ref{app:filament_nulls}, \ref{sec:naive_mocks}, \ref{sec:residualbias} and \ref{sec:error}.

\subsection{Data selection}\label{app:data}

We start with the DES Y3 shape catalogue \cite{DES:2020ekd}, applying all weak-lensing quality cuts used by the DES Collaboration \cite{DES:2020ekd}, together with the high-purity galaxy criteria from the DES Y3 GOLD sample \cite{DES:2020aks}. In addition, we impose a threshold on the position-angle (PA) uncertainty, $\sigma_\theta < 5^{\circ}$, to avoid possible residual PSF bias in galaxies with poorly constrained ellipticities and other artefacts that render the PA fit unstable. These measurements use the \texttt{metacalibration} (MC) algorithm \cite{DES:2020ekd}, with the unsheared ellipticity components corrected for PSF anisotropy with high precision. We tested an additional galaxy-by-galaxy PSF deconvolution to remove potential MC residuals and found that it changes the results by less than 0.3\%. Therefore, only the well-established MC measurements were adopted.

A shallow decision tree trained with \texttt{scikit-learn} \cite{scikit-learn}, based on the visual inspection of 10\,000 object images, was used to derive simple piecewise thresholds to separate artefacts and blends from galaxies. Accordingly, we perform an extra filtering, retaining objects that satisfy either $47.3<{\rm SNR}\leq75.2$ and $T_{\rm GAL}/T_{\rm PSF}\leq4.1$, or ${\rm SNR}>75.2$ and $T_{\rm GAL}/T_{\rm PSF}\leq15.7$. We found this to be a simple and effective quality cut to reduce galaxy misclassification, mainly by removing high-SNR artefacts and blended objects.

We divide this catalogue into two samples, disk-dominated galaxies and bulge-dominated galaxies, because the operative IA mechanisms are different. See Fig.~\ref{fig:ds_bs_density_map} for the \texttt{HEALPix}~\cite{Gorski:2004by} density maps of each tracer sample, and Fig.~\ref{fig:z_distribution} for their redshift distributions. Using the visually inspected objects, we also compare the PA measured visually, with \textsc{Morfometryka} \cite{Morfometryka}, and with MC, to verify the PA reference frame and the compatibility between independent PA estimates. Supplementary Figure~\ref{fig:delta_theta} shows the deviations between these PA measurements.

For the disk-dominated sample (DD), we consider only galaxies with robust late-type classifications by both \citet{DES:2020tkt} and \citet{2021MNRAS.507.4425C}, bulge-to-total fraction ${\rm B/T}\leq0.5$, axis ratio $b/a \le 0.766$, elongation $a/b<10$ to avoid thin streak artefacts, size ratio $T_{\rm GAL}/T_{\rm PSF}\ge 2$, and $0<z\leq0.9$. This yields a final DD sample of 3\,007\,638 galaxies. The $b/a$ cut removes nearly round objects for which the PA is ill-defined and noise-dominated, reducing PA scatter. The $a/b$ veto excludes pathologically elongated detections, typically trails, blended objects and other artefacts. These filters operate only on scalar elongation measures and are therefore PA-agnostic. The $T_{\rm GAL}/T_{\rm PSF}\ge 2$ requirement ensures that the galaxy is sufficiently resolved so that the PA is not driven by PSF anisotropy or model priors. Because late-type IA is intrinsically weak, these stricter PA-quality cuts reduce noise without biasing the LAIA measurement.

For the bulge-dominated sample (BD), we consider galaxies robustly classified as early-type by either \citet{DES:2020tkt} or \citet{2021MNRAS.507.4425C}, with ${\rm B/T}>0.5$, $b/a\leq0.9$, $a/b<10$, and $0<z\leq0.9$. Since the IA signal for ellipticals is expected to be substantially stronger than for disks, the $b/a$ threshold can be relaxed to gain statistics without materially degrading PA fidelity. This gives a final BD sample of 1\,381\,049 galaxies.

All galaxy selections described above are built from rotation-invariant scalar quantities, such as axis ratio $b/a$, elongation $a/b$ and the resolution factor $T_{\rm GAL}/T_{\rm PSF}$. None of these quantities depends on the absolute PA on the sky, so the cuts cannot directly imprint a preferred LAIA direction. They only modulate the PA signal-to-noise ratio. Residual correlations of the final PA field with PSF and survey-condition variables are removed and tested separately in Methods~\ref{sec:residual_correlations}.

For the filament analysis, we use the SDSS DR16 filament catalogue of \citet{Duque:2021xgw} as our main sample. We split this catalogue into Galactic-north (FN DR16) and Galactic-south (FS DR16) subsets. As an independent consistency check, we also use the SDSS DR12 filament catalogue of \citet{Chen:2015bra}, which covers the Galactic north only (FN DR12).

The reconstructed filament spines are converted into fixed comoving segments of length $10\,h^{-1}{\rm Mpc}$. For each segment, the projected PA is computed from the local spine tangent directions after projection onto the segment tangent plane. We retain only segments supported by at least five native catalogue spine samples. This minimum-sampling cut removes 3.0\% of FN DR16 segments, 6.1\% of FS DR16 segments and 2.3\% of FN DR12 segments. The final segment samples contain 104\,517 FN DR16 segments, 36\,623 FS DR16 segments and 97\,598 FN DR12 segments.

This fixed-length representation has three advantages. First, it gives comparable statistical weight to different parts of the filament skeleton, rather than allowing long filaments or densely sampled catalogue regions to dominate the estimator. Second, it reduces sensitivity to small-scale wiggles and catalogue sampling fluctuations while preserving the large-scale orientation of the spine. Third, it provides a uniform segment-level object on which the LAIA estimator and the filament permutation nulls can be applied. The minimum-sampling cut is also PA-agnostic: it depends only on the number of native spine samples supporting a segment, and therefore cannot by itself imprint a preferred sky direction.

Unlike galaxy PAs, filament PAs are not image-level shape measurements and therefore do not require PSF-, resolution- or ellipticity-based quality cuts. Their dominant possible systematics are instead footprint effects, redshift selection, variable skeleton sampling and local orientation anisotropies induced by the reconstruction. These are assessed with the binned-permutation null catalogues described in Methods~\ref{app:filament_nulls}. The sampling stability of the fixed-length filament segments is shown in Supplementary Figure~\ref{fig:filament_stats}. The left panel shows the number of native catalogue spine samples contributing to each retained $10\,h^{-1}{\rm Mpc}$ segment, while the right panel shows the signed axial residual between each local spine tangent direction and the final segment PA. These diagnostics show that the fiducial segment PAs are supported by multiple local spine samples and are stable against small-scale sampling fluctuations.

\subsection{Axial alignment estimators}\label{app:laia}

We obtain the object's sightline unit vector $\hat{\mathbf n}$ on the celestial sphere from the equatorial coordinate system, RA ($\alpha$) and Dec ($\delta$):
\begin{align}
    \hat{\mathbf{n}} = 
    \begin{bmatrix}
    \cos(\delta)\cos(\alpha) \\
    \cos(\delta)\sin(\alpha) \\
    \sin(\delta)
    \end{bmatrix}.
\end{align}

In order to build a plane tangent to the celestial sphere whose centre coincides with the object's coordinates, we can use an orthonormal basis:
\begin{equation}
\beta=\left\{\hat{\bm{\delta}} =\frac{1}{\left\|\frac{\partial \hat{\mathbf{n}}}{\partial \delta}\right\|} \frac{\partial \hat{\mathbf{n}}}{\partial \delta} \;\; {,} \;\;
\hat{\bm{\alpha}} = \frac{1}{\left\|\frac{\partial \hat{\mathbf{n}}}{\partial \alpha}\right\|}\frac{\partial \hat{\mathbf{n}}}{\partial \alpha} =
\frac{1}{\cos\delta}\frac{\partial \hat{\mathbf{n}}}{\partial \alpha} \right\}.
\end{equation}
Therefore, we can describe any vector on the plane as a linear combination of the vectors in this basis. The galaxy image is a two-dimensional projection on the tangent plane. This means the semi-major-axis direction of an elliptical galaxy ($\hat{\mathbf{a}}$) or the semi-minor axis of a spiral galaxy ($\hat{\mathbf{b}}$) can be written as a linear combination of the orthonormal basis $\beta$:
\begin{align}\label{eq:semiaxis}
    \hat{\mathbf{a}}(\alpha, \delta, \theta) &= \sin\theta  \hat{\bm\delta}-\cos\theta  \hat{\bm\alpha}, \\
    \hat{\mathbf{b}}(\alpha, \delta, \theta) &= \sin\theta  \hat{\bm\alpha} + \cos\theta  \hat{\bm\delta}, \nonumber
\end{align} 
where $\theta$ denotes the angle of the galaxy semi-major axis from west to north ($\theta=0 \Rightarrow \hat{\mathbf{a}}=-\hat{\bm{\alpha}}$). For filament segments, we use the same tangent-plane convention: if $\theta$ is the projected position angle of the filament spine, its headless orientation vector is
\begin{equation}\label{eq:spineaxis}
    \hat{\mathbf f}(\alpha,\delta,\theta)
    =
    \sin\theta\,\hat{\bm\delta}
    -
    \cos\theta\,\hat{\bm\alpha}.
\end{equation}

We want to find an axial IA with respect to a direction $\hat{\mathbf d}$, and therefore insensitive to the sign of the projections. Since galaxy semi-axes and filament spines define axial, or headless, directions rather than polar vectors, we identify $\hat{\mathbf a}_i \equiv -\hat{\mathbf a}_i$, $\hat{\mathbf b}_i \equiv -\hat{\mathbf b}_i$ and $\hat{\mathbf f}_i \equiv -\hat{\mathbf f}_i$. Defining a generic tracer orientation
\begin{equation}\label{eq:uaxis}
    \hat{\mathbf u}_i^X =
    \begin{cases}
    \hat{\mathbf a}_i, & X=BD,\\
    \hat{\mathbf b}_i, & X=DD,\\
    \hat{\mathbf f}_i, & X=F,
    \end{cases}
\end{equation}
the LAIA of a given tracer set can be measured with an estimator $\hat{E}^{X}(\hat{\mathbf d})$ that is a weighted average of the squared projections with respect to $\hat{\mathbf d}$:
\begin{equation}\label{eq:appestimator}
\hat{E}^{X}(\hat{\mathbf d})
=
\sum_{i=1}^{N_X}
\frac{
w_i^X{\bigl(\hat{\mathbf d}\cdot\hat{\mathbf u}_i^X\bigr)^2}
}{
1-\bigl(\hat{\mathbf d}\cdot\hat{\mathbf n}_i\bigr)^2
},
\qquad X\in\{BD,DD,F\}.
\end{equation}
The term $1-\bigl(\hat{\mathbf d}\cdot\hat{\mathbf n}_i\bigr)^2$ comes from projecting $\hat{\mathbf d}$ onto the tangent plane defined by $\beta$ and renormalizing the projected vector.

For the galaxy samples, the weights are related to the PA uncertainty $\sigma_\theta$:
\begin{equation}\label{appw_error}
    w_i^X=\frac{(\sigma_{\theta,i}^{X})^{-2}}{\displaystyle\sum_{j=1}^{N_X}(\sigma_{\theta,j}^{X})^{-2}},
    \qquad X\in\{BD,DD\}.
\end{equation}
Here $\sigma_{\theta,i}^{X}$ is the uncertainty in the PA (Methods~\ref{app:theta_error}). For filaments, no per-segment PA weighting is applied; equivalently, we use uniform normalized weights $w_i^f=1/N_f$.

We measure $\hat{E}^{X}(\hat{\mathbf d})$, $X\in\{BD,DD,F\}$, for different $\hat{\mathbf d}$ on a \texttt{HEALPix} grid and refine the maximum using a gradient-ascent optimization algorithm. The direction in which $\hat{E}^{X}(\hat{\mathbf d})$ is maximum is the LAIA direction of tracer $X$, $\hat{\mathbf d}_{\rm IA}^{X}$. Because the estimator is axial, $\hat{\mathbf d}_{\rm IA}^{X}$ and $-\hat{\mathbf d}_{\rm IA}^{X}$ are physically equivalent.

As derived in Methods~\ref{app:estimator}, by considering a simple phenomenological model with a nearly homogeneous departure from random position angles, we can convert the estimator value into an effective alignment amplitude $\theta_{\rm IA}^{X}$ with respect to $\hat{\mathbf d}_{\rm IA}^{X}$:
\begin{equation}
2\theta_{\rm IA}^{X}+\sin\bigl(2\theta_{\rm IA}^{X}\bigr)=
\frac{2\pi}{\overline{C}_X}
\Bigl(\hat{E}^{X}(\hat{\mathbf d}_{\rm IA}^{X})-\frac{1}{2}\Bigr).
\end{equation}
This is a transcendental equation that must be solved numerically. Here
\begin{equation}
\overline{C}_X=\sum_{i=1}^{N_X}w_i^X\,C_i^X,
\end{equation}
with $C_i^X=e^{-2(\sigma_{\theta,i}^{X})^2}$ for galaxy tracers $X\in\{BD,DD\}$ and $C_i^f=1$ for filaments. For galaxies, $\theta_{\rm IA}^{X}$ is an effective mean displacement of the relevant projected semi-axis; for filaments, it is an effective alignment amplitude of the projected spine directions.

\subsection{From shear measurements to galaxy position angle}\label{app:theta_error}

For a galaxy whose two-dimensional brightness distribution can be approximated by
an elliptical profile, the position angle $\theta$ can be written as function of the ellipticity parameters $e_1$ and $e_2$ as:
\begin{equation}
    \theta 
    = \frac{1}{2}\,\arctan2(e_2,\, e_1),
\end{equation}
where $\arctan2(e_2,e_1)$ is the two-argument arctangent that returns values in the range $(-\pi, \pi]$. Let $\Sigma_{(e_1,e_2)}$ be the $2\times 2$ covariance matrix of the ellipticity parameters:
\begin{equation}
    \Sigma_{(e_1,e_2)} 
    = 
    \begin{pmatrix}
        \sigma_{e_1}^2 & \mathrm{Cov}(e_1,e_2)\\
        \mathrm{Cov}(e_1,e_2) & \sigma_{e_2}^2
    \end{pmatrix}.
\end{equation}
We wish to propagate this uncertainty of $\theta$. First note that the gradient of $\theta$ (with respect to $e_1$ and $e_2$) is
\begin{equation}
    \!\!\!\nabla \theta 
    =
    \!\begin{pmatrix}
    \tfrac{1}{2}\,\dfrac{\partial}{\partial e_1} \arctan2(e_2, e_1)\\[6pt]
    \tfrac{1}{2}\,\dfrac{\partial}{\partial e_2} \arctan2(e_2, e_1)
    \end{pmatrix}\!\!
    = \!\!
    \begin{pmatrix}
    \displaystyle -\frac{e_2}{2(e_1^2 + e_2^2)}\\[8pt]
    \displaystyle \frac{e_1}{2(e_1^2 + e_2^2)}
    \end{pmatrix}.
    \label{eq:phi_grad}
\end{equation}
Using standard error propagation for correlated variables, the variance of $\theta$ is given by
\begin{equation}
    \sigma_\theta^2 
    = 
    \nabla \theta^{\,T}\,\Sigma_{(e_1,e_2)} \,\nabla \theta,
\end{equation}
we obtain
\begin{align}\label{eq:sigma_theta_ap}
    \sigma_\theta^2 
    &= 
    \frac{\Bigl[
    \,e_2^2\,\sigma_{e_1}^2 
    \;+\;
    e_1^2\,\sigma_{e_2}^2
    \;-\;
    2\,e_1\,e_2 \,\mathrm{Cov}(e_1, e_2)
    \Bigr]}{4\,(e_1^2 + e_2^2)^2}
    .
\end{align} In practice, we impose a $1^\circ$ lower bound to stabilize the weighting ($\sigma_{\theta,i}\!\to\!\max(\sigma_{\theta,i},1^\circ)$, tests with $0.75^\circ$ and $0.5^\circ$ gave equivalent results). This choice reflects our focus on a global dipole estimate rather than dominance by a few high-SNR objects.

\subsection{How to compute $\theta_{\rm IA}$}\label{app:estimator}

We start assuming that the measurement error $\delta_i$ follows a Gaussian distribution with standard deviation $\sigma_{\theta,i}^{X}$ for galaxy tracers $X\in\{BD,DD\}$, where $\sigma_{\theta,i}^{X}$ denotes the PA error. Its probability density function is
\begin{equation}\label{eq:phi_error_dist}
    P_i(\delta) =
    \frac{1}{\sqrt{2\pi}\,\sigma_{\theta,i}^{X}}
    e^{-\delta^2/(2(\sigma_{\theta,i}^{X})^2)}.
\end{equation}
For filaments, $X=F$, we do not apply a PA-error attenuation and set the corresponding attenuation factor to unity, as described below.

We can now expand the scalar product of the estimator in terms of the angle between the projected trial direction and the tracer orientation. For compactness, we write the tracer orientation as $\hat{\mathbf u}_i^X$, with $\hat{\mathbf u}_i^{BD}=\hat{\mathbf a}_i$ for bulge-dominated galaxies, $\hat{\mathbf u}_i^{DD}=\hat{\mathbf b}_i$ for disk-dominated galaxies and $\hat{\mathbf u}_i^F=\hat{\mathbf f}_i$ for filament spines. Let $\phi_i$ be the angle between $\hat{\mathbf u}_i^X$ and the projected preferred direction. Then
\begin{align}
  {\bigl(\hat{\mathbf d}\cdot\hat{\mathbf u}_i^X\bigr)}^2 \nonumber
&=
\cos^2\bigl(\phi_i+\delta_i\bigr)
=
\frac{1}{2}
+\frac{1}{2}\cos\bigl(2(\phi_i+\delta_i)\bigr)
\\ &= 
\frac{1}{2}\bigl[\,1+\cos\bigl(2\phi_i\bigr)\,\cos\bigl(2\delta_i\bigr)-\sin\bigl(2\phi_i\bigr)\,\sin\bigl(2\delta_i\bigr)\bigr]. 
\end{align}
Such that the average is given by
\begin{align}
\bigl\langle{\bigl(\hat{\mathbf d}\cdot\hat{\mathbf u}_i^X\bigr)}^2\bigr\rangle
&=
\frac{1}{2}
+\frac{1}{2}\bigl\langle\cos\bigl(2\phi_i\bigr)\,\cos\bigl(2\delta_i\bigr)\bigr\rangle_{\phi,\delta} \nonumber \\
&=
\frac{1}{2}
+\frac{C_i^X}{2}\,\bigl\langle\cos\bigl(2\phi_i\bigr)\bigr\rangle_{\phi}.
\end{align}
Here $C_i^X$ measures the attenuation of the alignment signal due to PA uncertainty. For galaxy tracers it is computed as an average of $\cos(2\delta_i)$ over the Gaussian error distribution:
\begin{equation}
C_i^X = \langle \cos 2\delta_i \rangle_\delta
= \int_{-\infty}^\infty \cos(2\delta) \, P_i(\delta) \, \mathrm{d}\delta
= e^{-2(\sigma_{\theta,i}^{X})^2}.
\end{equation}
Thus, $C_i^X$ varies from 1, for perfectly preserved alignment, to 0, when the signal is fully washed out by PA noise. For filaments we set
\begin{equation}
C_i^f=1,
\end{equation}
corresponding to no PA-error attenuation.

We should also consider the ``saturation'' effect of the average alignment introduced by LAIA. In other words, if the effective alignment amplitude is $\theta_{\rm IA}^{X}$, when a tracer has $\phi_i<\theta_{\rm IA}^{X}$, it is displaced by less than the average amount to reach the alignment direction. Considering a Universe without LAIA, a tracer would have a random angular distance $\phi_i$ from the preferred axis; with LAIA, it has $\phi_i^{\prime}$ given by
\begin{equation}
\phi_i^{\prime}
=
\begin{cases}\label{eq:saturation}
0,&0\le\phi<\theta_{\rm IA}^{X},\\
\phi-\theta_{\rm IA}^{X},&\theta_{\rm IA}^{X}\le\phi\le\frac{\pi}{2}.
\end{cases}
\end{equation}
Assuming a homogeneous distribution of angular distances without LAIA, we have $K(\phi)=2/\pi$. With LAIA:
\begin{align}
\bigl\langle\cos\bigl(2\phi_i'\bigr)\bigr\rangle
&=
\int_{0}^{\pi/2}\cos\bigl(2\phi_i'\bigr)\,K(\phi)\,d\phi
\nonumber\\ &= 
\frac{2}{\pi}\Bigl[\int_{0}^{\theta_{\rm IA}^{X}}1\,d\phi
+\int_{\theta_{\rm IA}^{X}}^{\pi/2}\cos\bigl(2\phi-2\theta_{\rm IA}^{X}\bigr)\,d\phi\Bigr]
\nonumber\\&=
\frac{2}{\pi}\Bigl[\theta_{\rm IA}^{X}+\frac{1}{2}\sin\bigl(2\theta_{\rm IA}^{X}\bigr)\Bigr].   
\end{align}
Therefore,
\begin{equation}
\bigl\langle{\bigl(\hat{\mathbf d}\cdot\hat{\mathbf u}_i^X\bigr)}^2\bigr\rangle
=
\frac{1}{2}
+\frac{C_i^X}{\pi}\Bigl[\theta_{\rm IA}^{X}+\frac{1}{2}\sin\bigl(2\theta_{\rm IA}^{X}\bigr)\Bigr].
\end{equation}
As a result, we can estimate the effective alignment amplitude of tracer $X$ with respect to $\hat{\mathbf d}_{\rm IA}^{X}$ through
\begin{equation}
\hat{E}^{X}(\hat{\mathbf d}_{\rm IA}^{X})
=
\frac{1}{2}
+\frac{\overline{C}_X}{\pi}\Bigl[\theta_{\rm IA}^{X}+\frac{1}{2}\sin\bigl(2\theta_{\rm IA}^{X}\bigr)\Bigr],
\end{equation}
where
\begin{equation}
\overline{C}_X
=
\sum_{i=1}^{N_X}w_i^X\,C_i^X.
\end{equation}
For galaxy tracers, $C_i^X=e^{-2(\sigma_{\theta,i}^{X})^2}$, while for filaments $C_i^f=1$. The constant $1/2$ is the isotropic baseline of the squared-projection statistic: since $\cos^2\phi=(1+\cos2\phi)/2$, uniformly distributed headless PAs have $\langle\cos^2\phi\rangle=1/2$, and $\hat{E}^X-1/2$ isolates the coherent axial $\cos2\phi$ excess. This motivates the quadratic projection rather than an absolute-projection statistic. Finally, the equation for $\theta_{\rm IA}^{X}$ is
\begin{equation}
2\theta_{\rm IA}^{X}+\sin\bigl(2\theta_{\rm IA}^{X}\bigr)
=
\frac{2\pi}{\overline{C}_X}
\Bigl(\hat{E}^{X}(\hat{\mathbf d}_{\rm IA}^{X})-\frac{1}{2}\Bigr),
\end{equation}
which is a transcendental equation. 

\subsection{Lensing leakage}\label{sec:lensing}

In standard weak-lensing analyses, intrinsic alignments are usually treated as a contaminant of the lensing shear signal. Here we consider the converse question: whether lensing or kinematic effects could contaminate a sky-scale IA measurement. Kinematic aberration from our motion changes apparent positions and number counts, but leaves galaxy shapes unchanged at leading order~\cite{Calvao:2004ma}. It therefore does not generate a net sky-scale axial alignment that could mimic LAIA. Weak-lensing shear is a spin-2, locally quadrupolar distortion; in an ideal statistically isotropic full-sky survey, it would not generate a net sky-scale axial alignment that could mimic LAIA. Lensing magnification can modulate sizes, number counts and effective sample weights across the footprint, but it does not rotate position angles~\cite{Bartelmann:1999yn}. Consequently, LAIA should be insensitive to kinematic aberration and only weakly affected by lensing in the absence of survey anisotropies.

In real data, however, lensing can couple to the survey footprint, selection function and spatially varying weights. Shear can perturb measured galaxy position angles, while magnification can modulate which galaxies enter the sample and how they are weighted. Through an incomplete and anisotropic mask, these coherent but weak perturbations may not cancel exactly and could induce a small shift in the recovered LAIA axis or amplitude. We refer to this footprint-mediated coupling as lensing leakage.

To assess this effect, we use the BUZZARD $N$-body lightcone with ray tracing, selecting 55\,million galaxies in the same redshift range as the DES Y3 galaxy samples and applying the DES footprint, rotated to the simulation quadrant for consistency. We run the alignment pipeline twice, once with lensing applied to the measured position angles and once without. The recovered $\theta_{\rm IA}$ differs by $1' \pm 0.2'$.

Although this is a small effect, we propagate it conservatively in the galaxy measurements. We model the lensing--footprint coupling as a stochastic axial rotation: the amplitude is fixed by the BUZZARD calibration, while the direction is unknown a priori. Accordingly, in each simulation or bootstrap resampling we apply an additional axial rotation of $1'$ in a random isotropically distributed direction. This procedure inflates the error bars to capture the realistic, footprint-mediated impact of lensing on the galaxy-based LAIA estimator. The filament measurements are not image-level galaxy-shape measurements and therefore do not require this lensing-leakage correction.

\subsection{$N$-body simulation-based realistic null samples}
\label{app:nbody_sampling}

\begin{figure*}[t] 
    \includegraphics[width=0.9\textwidth]{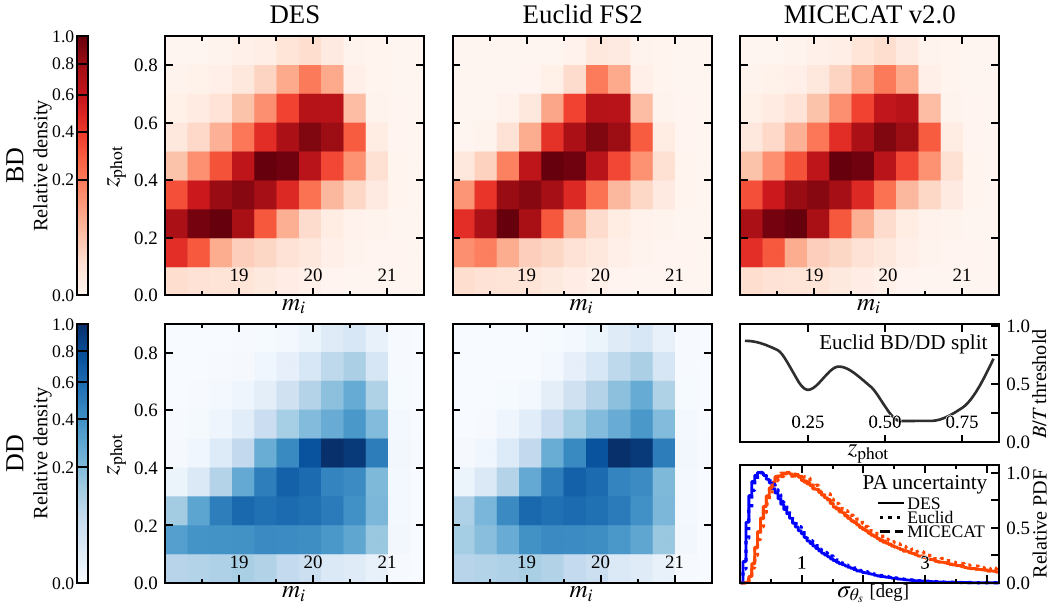}
\caption{
\textbf{Construction and validation of the simulation-based realistic null samples.}
Joint distributions of photometric redshift, $z_{\rm phot}$, and corrected $i$-band magnitude, $m_i$, for the DES reference samples and the corresponding Euclid FS2 and MICECAT v2.0 mock samples. The upper row shows the bulge-dominated (BD) sample and the lower row shows the disc-dominated (DD) sample.
Colour encodes the relative density in each $(z_{\rm phot},m_i)$ cell, normalized by the peak cell density within each panel. The DES BD and DD samples define the target two-dimensional selection functions, which are matched by down-sampling the Euclid FS2 catalogues after applying a redshift-dependent bulge-to-total threshold. MICECAT v2.0 contributes only the red/IA-enabled population and is therefore shown only for the BD-like case. The lower-right upper inset shows the smooth redshift-dependent Euclid $B/T$ threshold used to split the simulation into BD and DD populations. The lower-right lower inset compares the position-angle uncertainty distributions, $\sigma_{\theta_s}$, in DES with the uncertainties assigned to the Euclid FS2 and MICECAT v2.0 matched samples. Colours distinguish BD and DD samples, while line style distinguishes the data source: solid for DES, dotted for Euclid FS2 and dashed for MICECAT v2.0. This validation shows that the simulation-based null catalogues reproduce the relevant DES selection in redshift, magnitude and angular measurement uncertainty before being used to estimate the intrinsic-alignment null distributions.
}
    \label{fig:nbody_sampling}
\end{figure*}

We construct simulation-based realistic null catalogues from two independent $N$-body-based mock data sets, MICECAT v2.0 and the Euclid FS2 light-cone. Both mocks are based on state-of-the-art cosmological simulations that include standard intrinsic-alignment information within an isotropic $\Lambda$CDM framework. They therefore provide a controlled way to test the IA estimator on catalogues in which galaxy orientations arise from conventional large-scale-structure physics, while the observational selection is matched to that of the DES samples. The purpose of this procedure is not to reproduce the measured DES signal, but to generate realistic null samples with the same selection-dependent quantities entering the estimator. In particular, we match the joint distribution of photometric redshift and corrected $i$-band magnitude, $(z_{\rm phot},m_i)$, and assign position-angle uncertainties, $\sigma_{\theta_s}$, drawn from the corresponding DES distributions. The final matched samples are shown in Extended Data Figure~\ref{fig:nbody_sampling}.

The MICECAT v2.0 IA catalogue provides a particularly direct construction. We use the red-galaxy IA population, which corresponds to the elliptical, bulge-dominated (BD) component relevant for the IA model. We therefore construct a BD-like MICECAT realistic null sample and do not build a disc-dominated (DD) MICECAT analogue, since the corresponding DD IA population is not available. The MICECAT red galaxies are down-sampled to match the DES BD selection in the two-dimensional $(z_{\rm phot},m_i)$ plane. We define a DES target histogram
\begin{equation}
    H^{\rm DES,BD}_{jk}
    =
    N^{\rm DES,BD}(z_j,m_k),
\end{equation}
using bins of width $\Delta z=0.1$ and $\Delta m_i=0.3$, and select MICECAT objects in each cell without replacement. The number of selected objects is
\begin{equation}
    H^{\rm MICE,out}_{jk}
    =
    \min\!\left[
        H^{\rm DES,BD}_{jk},
        H^{\rm MICE,red}_{jk}
    \right].
\end{equation}
The overlap between the DES BD selection and the MICECAT red population is very large: the final matched MICECAT sample is reduced by only $\simeq 0.3\%$ relative to the DES target. This small loss has negligible impact on the effective statistical power of the mock null sample.

For the Euclid FS2 catalogue, an additional step is required because the mock does not contain the same BD/DD classification used for the DES galaxies. It does, however, provide a bulge-to-total quantity, denoted here by $B/T$. We use this information to define a redshift-dependent threshold, $t(z)$, chosen to reproduce the DES BD fraction in each redshift bin. Specifically, in a redshift bin $j$ we compute
\begin{equation}
    f_{\rm BD}^{\rm DES}(z_j)
    =
    \frac{N_{\rm BD}^{\rm DES}(z_j)}
    {N_{\rm BD}^{\rm DES}(z_j)+N_{\rm DD}^{\rm DES}(z_j)} ,
\end{equation}
and choose the Euclid threshold $t_j$ such that
\begin{equation}
    {\rm Pr}_{\rm Euclid}\!\left(B/T \geq t_j \mid z_j\right)
    \simeq
    f_{\rm BD}^{\rm DES}(z_j).
\end{equation}
Objects with $B/T \geq t_j$ are assigned to the Euclid BD mock sample, and the remaining objects in the same redshift bin are assigned to the Euclid DD mock sample. This avoids imposing a single global morphology threshold and instead matches the redshift evolution of the DES BD/DD composition.

After this split, the Euclid BD and DD catalogues are independently matched to the corresponding DES two-dimensional selection functions in $(z_{\rm phot},m_i)$. For each class $S\in\{\mathrm{BD},\mathrm{DD}\}$, we define
\begin{equation}
    H^{\rm DES}_{S,jk}
    =
    N^{\rm DES}_{S}(z_j,m_k),
\end{equation}
and select Euclid galaxies without replacement in each cell:
\begin{equation}
    H^{\rm Euclid,out}_{S,jk}
    =
    \min\!\left[
        H^{\rm DES}_{S,jk},
        H^{\rm Euclid}_{S,jk}
    \right].
\end{equation}

This ``allow-shortfall'' prescription preserves the DES selection wherever the simulation has support, avoids duplicating mock galaxies, and prevents a small number of sparsely populated edge cells from forcing a global down-scaling of the entire sample. The matched Euclid DD sample loses only $\simeq 3.9\%$ of the DES target number of galaxies, corresponding to an expected reduction of less than $\simeq 2\%$ in the effective signal-to-noise ratio. The Euclid BD sample has a larger shortfall, with a loss of about $10\%$ of the DES target number of galaxies. This corresponds to an expected effective signal-to-noise reduction of approximately $3.5\%$ in our weighted estimator, and we keep this sample as a conservative simulation-based null.

The position-angle uncertainty $\sigma_{\theta_s}$ (see Equation~\ref{eq:sigma_theta_ap}) assigned to the simulation galaxies is matched to DES in a consistent way. For both MICECAT and Euclid FS2, $\sigma_{\theta_s}$ is drawn from the DES observed sample distribution conditional on the same $(z_{\rm phot},m_i)$ cell used for the abundance matching, considering each morphological classification.

We do not perturb the simulated position angles by these assigned uncertainties. The simulated orientation, through $\theta_s$, defines the galaxy direction entering the estimator, while $\sigma_{\theta_s}$ enters only as the measurement uncertainty used in the weighting and normalization of the estimator. This treatment keeps the simulation orientations fixed (maintaining IA prescriptions intact) and matches the DES observational uncertainty model, making the samples directly comparable to the other realistic null catalogues used in the analysis.

Extended Data Figure~\ref{fig:nbody_sampling} summarizes the outcome of this construction. The DES, Euclid FS2 and MICECAT v2.0 panels show that the matched catalogues reproduce the target two-dimensional selection in $(z_{\rm phot},m_i)$ for the relevant BD and DD samples. The Euclid inset shows the redshift-dependent $B/T$ threshold used to construct the BD/DD split, and the PA-uncertainty inset shows that the assigned mock uncertainties follow the corresponding DES distributions. These DES-matched simulation catalogues are then used as $N$-body-based realistic nulls for the IA estimator.

\subsection{\label{sec:residual_correlations}Residual spurious correlations removal and galaxy empirical nulls}

A robust measurement of a footprint-scale axial-alignment signal requires the observed position-angle field to be insensitive to PSF, instrumental and survey-condition systematics. Spurious correlations can arise through several routes, including direct PSF leakage into the measured galaxy shapes, spatially varying observing conditions, foreground extinction, depth variations, or selection effects that couple the retained galaxy sample to image quality. The cut on the position-angle uncertainty, $\sigma_\theta$, is one possible route through which such effects could enter our sample, since galaxies with well-measured position angles may occupy a restricted range of size, morphology or signal-to-noise ratio. These properties can in turn correlate with the PSF or with local observing conditions. Our goal is to remove any component of the observed spin-2 orientation field that is spuriously correlated with PSF or survey-condition diagnostics. We therefore apply an explicit residualisation step to the final selected catalogues, and use the residualised catalogue for the LAIA measurements reported in the main text.

For each galaxy \(i\), we define the observed spin-2 ellipticity/orientation vector as
\begin{equation}
    \mathbf e_{{\rm obs},i}
    \equiv
    \left(e_{{\rm obs},1,i},e_{{\rm obs},2,i}\right),
\end{equation}
where \(j=1,2\) labels the two ellipticity components in the same local tangent-plane convention used by the estimator. We model the observed field ($\mathbf e_{{\rm obs},i}$) as the sum of different components,
\begin{equation}   
    \mathbf e_{{\rm obs},i}
    \equiv
    \hat{\mathbf e}_{{\rm sys},i}
    +
    \mathbf e_{{\rm phys},i} + \mathbf e_{{\rm noise},i} \;, 
\end{equation}
or in a different decomposition,
\begin{equation}
    \mathbf e_{{\rm obs},i}
    \equiv
    \hat{\mathbf e}_{{\rm sys},i}
    +
    \mathbf e_{{\rm res},i} \;.
    \label{eq:obs_sys_res_decomposition}    
\end{equation}
Here \(\hat{\mathbf e}_{{\rm sys},i}\) denotes the component of the observed orientation field that is linearly associated with PSF and survey-conditions. It is not interpreted as a physical galaxy signal. The residual field, \(\mathbf e_{{\rm res},i}\), is the part of the measured orientation field left after subtracting this fitted systematic component. In this operational sense, the residual field is the component treated as carrying the intrinsic or physical orientation signal of interest ($\mathbf e_{{\rm phys},i}$), together with the remaining measurement noise ($\mathbf e_{{\rm noise},i}$).

The diagnostic vector for galaxy \(i\), \(\mathbf x_i\), contains the PSF ellipticity components, the PSF size, the observing-condition quantities used in the diagnostic panels, and first-order interaction terms between PSF and survey-condition quantities:
\begin{equation}
\begin{split}
    \mathbf x_i =
    \big(
    &p_{1,i},p_{2,i},T_{{\rm PSF},i},
    {\rm FWHM}_i,A_i,m_{{\rm lim},i},  \\
    &\sqrt{{\rm skyvar}_i},E(B-V)_i,
    \{s_{\ell,i}p_{m,i}\}_{\ell,m}
    \big).
\end{split}
\label{eq:esys_design_vector}
\end{equation}
In this expression, \(p_{1,i}\) and \(p_{2,i}\) are the two PSF ellipticity components, \(T_{{\rm PSF},i}\) is the PSF size, \({\rm FWHM}_i\) is the image full width at half maximum, \(A_i\) is the airmass, \(m_{{\rm lim},i}\) is the limiting magnitude, \(\sqrt{{\rm skyvar}_i}\) is the square root of the local sky brightness variance, and \(E(B-V)_i\) is the Galactic reddening. The index \(\ell\) runs over the survey-condition variables,
\begin{equation}
    s_{\ell,i}
    \in
    \left\{
    {\rm FWHM}_i,
    A_i,
    m_{{\rm lim},i},
    \sqrt{{\rm skyvar}_i},
    E(B-V)_i
    \right\},
\end{equation}
while the index \(m\) runs over the PSF quantities,
\begin{equation}
    p_{m,i}
    \in
    \left\{
    p_{1,i},
    p_{2,i},
    T_{{\rm PSF},i}
    \right\}.
\end{equation}
The interaction terms, \(\{s_{\ell,i}p_{m,i}\}_{\ell,m}\), allow PSF leakage to depend on observing conditions while keeping the model linear in its fitted parameters.

To reduce the risk of fitting coherent modes in the same sky region where the correction is evaluated, the regression is performed as a spatial cross-fit. The catalogue is first assigned to coarse HEALPix blocks at \(N_{\rm side}=16\), corresponding to pixels of area \(4\pi/[12(16)^2]\simeq13.4\,{\rm deg}^2\), or a characteristic angular scale of \(\simeq3.7^\circ\). The unique \(N_{\rm side}=16\) pixels are then split into five folds with approximately equal numbers of coarse pixels, and each galaxy inherits the fold of its parent pixel. For a test fold \(\mathcal F_f\), the training set is the complement \(\mathcal T_f\). All standardisation and coefficient fitting are performed only on \(\mathcal T_f\), and the fitted model is then evaluated out of fold on galaxies in \(\mathcal F_f\).

For each fold \(f\), and for each ellipticity component \(j=1,2\), the columns of the design matrix are standardised using the training set,
\begin{equation}
    \tilde x^{(f)}_{ik}
    =
    \frac{x_{ik}-\mu_{f,k}}{s_{f,k}},
    \qquad i\in\mathcal T_f .
\end{equation}
Here \(k=1,\ldots,K\) labels the \(K\) columns of the diagnostic vector \(\mathbf x_i\), including the interaction terms. The quantities \(\mu_{f,k}\) and \(s_{f,k}\) are the mean and standard deviation of diagnostic column \(k\), computed using only the training galaxies in \(\mathcal T_f\). The same training-set mean and standard deviation are then used to standardise the corresponding diagnostic values for the held-out galaxies in \(\mathcal F_f\).

The systematic component is fitted independently for the two ellipticity components using an ordinary multilinear regression. For component \(j\), the model is
\begin{equation}
    e_{{\rm obs},j,i}
    =
    \alpha_{f,j}
    +
    \sum_{k=1}^{K}
    \beta_{f,jk}\tilde x^{(f)}_{ik}
    +
    \epsilon_{j,i},
    \qquad i\in\mathcal T_f .
    \label{eq:esys_linear_model}
\end{equation}
This is a simultaneous linear fit over all diagnostic dimensions: \(\alpha_{f,j}\) is a constant offset for fold \(f\) and component \(j\), while each coefficient \(\beta_{f,jk}\) gives the response of that ellipticity component to one standardised diagnostic column. Equivalently, the fitted coefficients are obtained from
\begin{equation}
    \left(\hat\alpha_{f,j},\hat{\boldsymbol\beta}_{f,j}\right)
    =
    \arg\min_{\alpha,\boldsymbol\beta}
    \sum_{i\in\mathcal T_f}
    \left[
        e_{{\rm obs},j,i}
        -
        \alpha
        -
        \sum_{k=1}^{K}\beta_k\tilde x^{(f)}_{ik}
    \right]^2 .
    \label{eq:esys_ols_fit}
\end{equation}
The out-of-fold systematic prediction for a galaxy \(i\in\mathcal F_f\) is then
\begin{equation}
    \hat e_{{\rm sys},j,i}
    =
    \hat\alpha_{f,j}
    +
    \sum_{k=1}^{K}
    \hat\beta_{f,jk}\tilde x^{(f)}_{ik},
    \qquad j=1,2 .
    \label{eq:esys_prediction}
\end{equation}
Finally, the residual field used by the estimator is defined by subtracting the fitted systematic component from the observed field,
\begin{equation}
    e_{{\rm res},j,i}
    =
    e_{{\rm obs},j,i}
    -
    \hat e_{{\rm sys},j,i}.
    \label{eq:esys_residual}
\end{equation}
This type of mitigation is standard in weak-lensing analyses, where additive shape systematics are routinely modelled and tested as functions of PSF quantities and survey properties \citep{DES:2020ekd,DES:2021gua}. The corresponding residual position angle is obtained from the residual spin-2 vector,
\begin{equation}
    \theta_{{\rm res},i}
    =
    \frac{1}{2}
    \arg\left(e_{{\rm res},1,i}+{\rm i}e_{{\rm res},2,i}\right),
\end{equation}
with the same headless-angle convention used throughout the analysis.

We validate the correction by measuring trends of the mean ellipticity components with each PSF and survey-condition diagnostic. For a diagnostic \(q\), galaxies are sorted into decile bins of \(q\), and each bin is plotted at the mean value of \(q\) within that bin. In bin \(b\), we compute
\begin{equation}
    \left\langle e_{X,j}\right\rangle_b
    =
    \frac{1}{N_b}
    \sum_{i\in b} e_{X,j,i},
    \qquad
    X\in\{{\rm obs},{\rm sys},{\rm res}\}.
\end{equation}
The same binning procedure is applied to the observed field, the fitted systematic component and the residual field.

\begin{figure*}[p]
    \centering
    \includegraphics[width=0.869\textwidth]{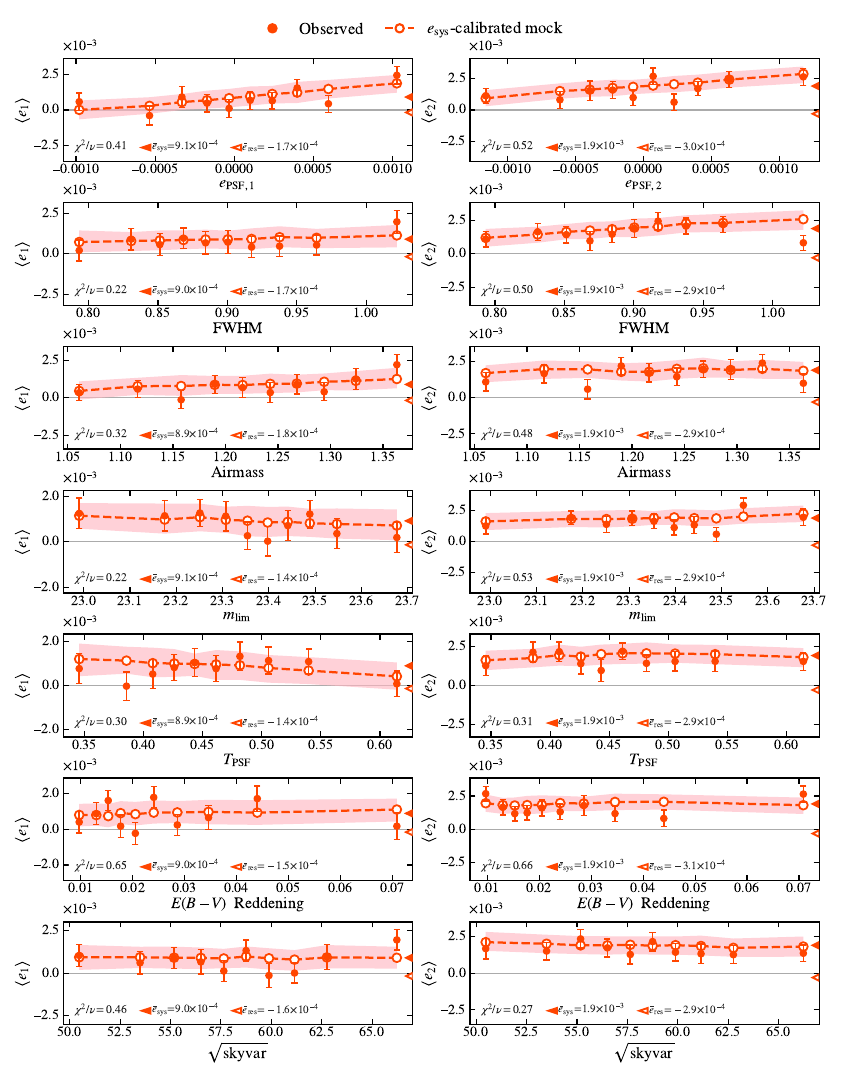}
\caption{\textbf{Raw shape--diagnostic trends for the bulge-dominated sample.}
Mean observed ellipticity components of bulge-dominated (BD) galaxies are shown as a function of the PSF, observing-condition and foreground diagnostics used in the systematic-error assessment. Measurements are binned in deciles of the corresponding diagnostic variable, with each point placed at the mean value within the bin. The two columns correspond to $\langle e_1\rangle$ and $\langle e_2\rangle$, and the rows show, from top to bottom, $e_{\rm PSF}$, FWHM, airmass, limiting magnitude $m_{\rm lim}$, PSF size $T_{\rm PSF}$, Galactic reddening $E(B-V)$ and $\sqrt{\rm skyvar}$. Filled points with error bars show the observed raw measurements, while open points and dashed curves show the $e_{\rm sys}$-calibrated realistic-mock expectation; shaded bands enclose the central 68 per cent mock interval. The in-panel values give the reduced $\chi^2/\nu$ for the data--mock comparison, together with the mean systematic and residual ellipticity offsets, $e_{\rm sys}$ and $e_{\rm res}$. Filled and open triangles on the right-hand axis indicate $e_{\rm sys}$ and $e_{\rm res}$, respectively.}
    \label{fig:bulge_e_sys_raw}
\end{figure*}

\begin{figure*}[p]
    \centering
    \includegraphics[width=0.869\textwidth]{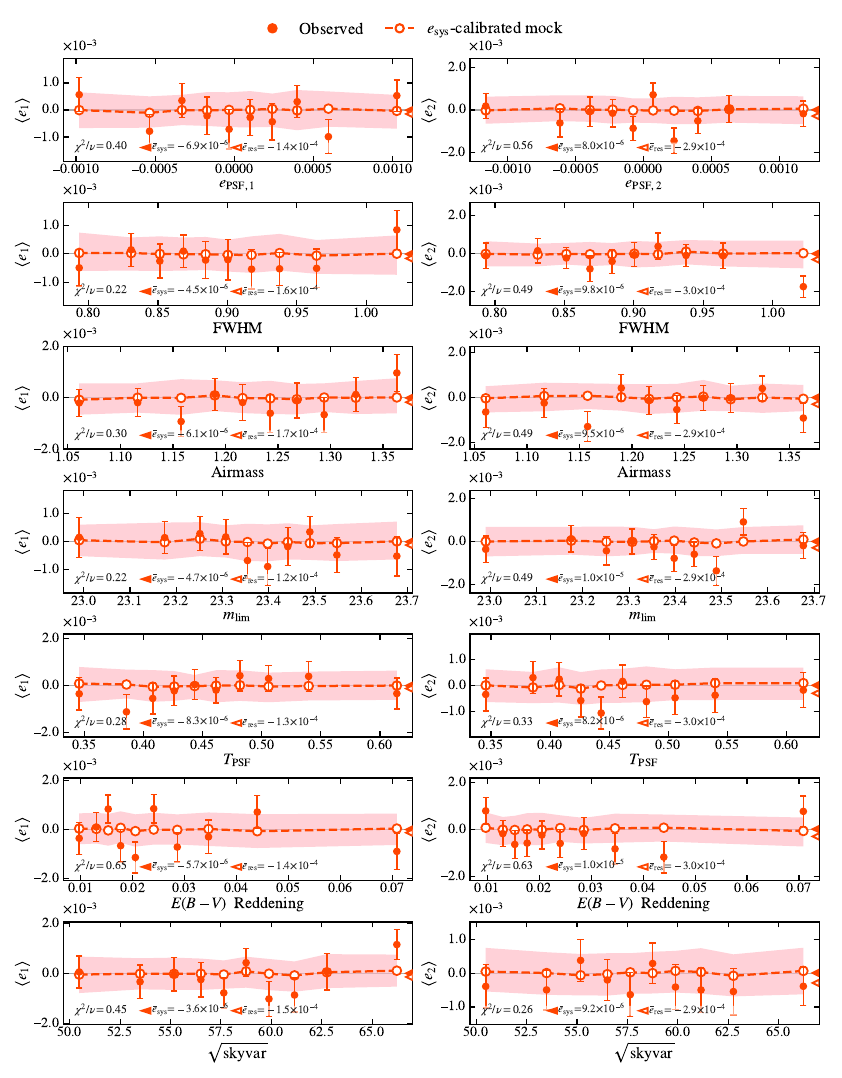}
\caption{\textbf{Residual shape--diagnostic trends for the bulge-dominated sample.}
Residual ellipticity trends for the bulge-dominated (BD) sample are shown after subtraction of the systematic component. Measurements are binned in deciles of the corresponding diagnostic variable, with each point placed at the mean value within the bin. The figure uses the same layout and diagnostics as Extended Data Figure~\ref{fig:bulge_e_sys_raw}: the left and right columns show $\langle e_1\rangle$ and $\langle e_2\rangle$, while rows correspond to $e_{\rm PSF}$, FWHM, airmass, limiting magnitude $m_{\rm lim}$, PSF size $T_{\rm PSF}$, Galactic reddening $E(B-V)$ and $\sqrt{\rm skyvar}$. Filled points with error bars show the observed residual measurements, open points and dashed curves show the residual-null expectation, and shaded regions indicate the central 68 per cent interval of the residual-null distribution. The in-panel annotations give the reduced $\chi^2/\nu$, $e_{\rm sys}$ and $e_{\rm res}$; filled and open triangles on the right-hand axis mark the systematic and residual mean offsets, respectively.}
    \label{fig:bulge_e_sys_residual}
\end{figure*}

\begin{figure*}[p]
    \centering
    \includegraphics[width=0.869\textwidth]{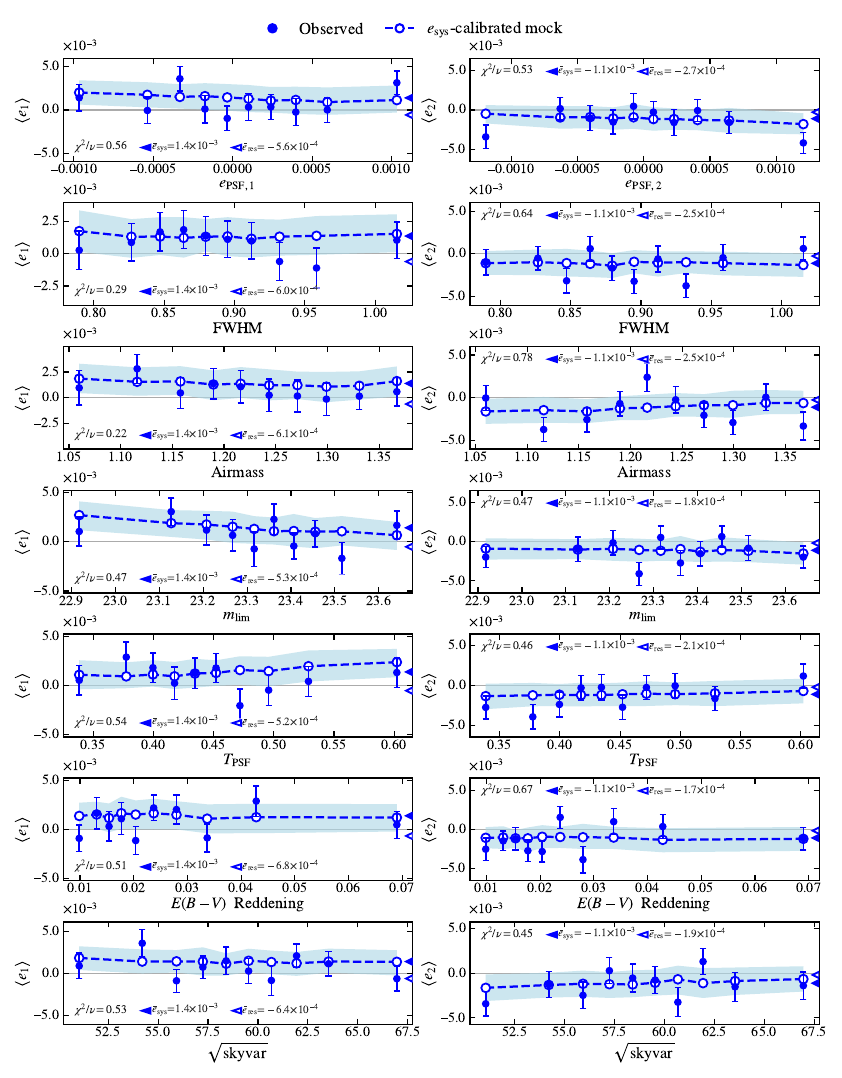}
\caption{\textbf{Raw shape--diagnostic trends for the disc-dominated sample.}
Mean observed ellipticity components of disc-dominated (DD) galaxies are shown as a function of the PSF, observing-condition and foreground quantities used in the systematic-error assessment. Measurements are binned in deciles of the corresponding diagnostic variable, with each point placed at the mean value within the bin. The left and right columns show $\langle e_1\rangle$ and $\langle e_2\rangle$, respectively, while rows correspond to $e_{\rm PSF}$, FWHM, airmass, limiting magnitude $m_{\rm lim}$, PSF size $T_{\rm PSF}$, Galactic reddening $E(B-V)$ and $\sqrt{\rm skyvar}$. Filled circles with error bars denote the observed measurements, and open circles connected by dashed curves show the corresponding $e_{\rm sys}$-calibrated realistic-mock expectation. Shaded regions indicate the central 68 per cent interval of the mock distribution. In each panel, the annotations report the reduced $\chi^2/\nu$ of the data--mock comparison and the mean systematic and residual ellipticity offsets, $e_{\rm sys}$ and $e_{\rm res}$; filled and open triangles on the right-hand axis mark these two offsets, respectively.}
    \label{fig:disc_e_sys_raw}
\end{figure*}

\begin{figure*}[p]
    \centering
    \includegraphics[width=0.869\textwidth]{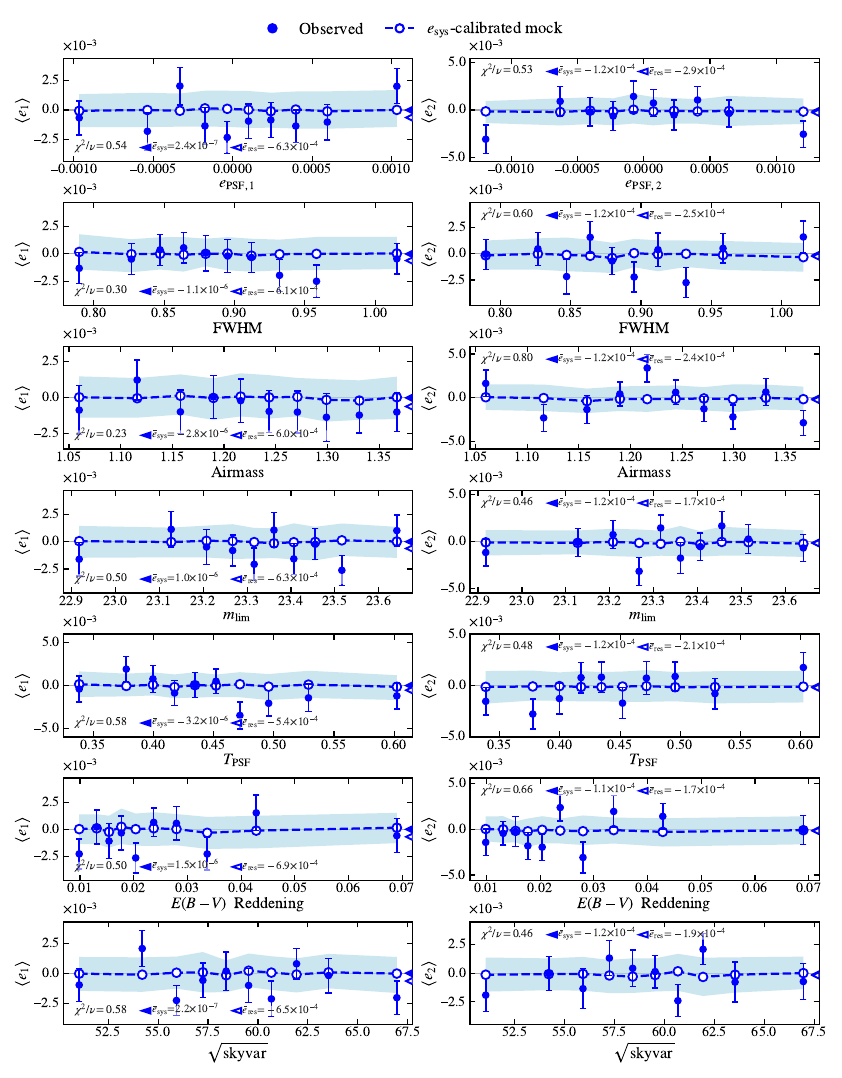}
\caption{\textbf{Residual shape--diagnostic trends for the disc-dominated sample.}
Residual ellipticity trends for the disc-dominated (DD) sample are shown after subtraction of the systematic component. Measurements are binned in deciles of the corresponding diagnostic variable, with each point placed at the mean value within the bin. The layout follows Extended Data Figure~\ref{fig:disc_e_sys_raw}: columns show $\langle e_1\rangle$ and $\langle e_2\rangle$, and rows show $e_{\rm PSF}$, FWHM, airmass, limiting magnitude $m_{\rm lim}$, PSF size $T_{\rm PSF}$, Galactic reddening $E(B-V)$ and $\sqrt{\rm skyvar}$. Filled circles with error bars denote the observed residual signal, and open circles connected by dashed curves show the corresponding residual-null expectation; shaded regions show the central 68 per cent interval of the residual-null distribution. The annotations report the reduced $\chi^2/\nu$ of the residual comparison and the mean systematic and residual offsets, $e_{\rm sys}$ and $e_{\rm res}$, with filled and open triangles marking these quantities on the right-hand axis.}
    \label{fig:disc_e_sys_residual}
\end{figure*}

Extended Data Figures~\ref{fig:bulge_e_sys_raw} and \ref{fig:disc_e_sys_raw} show the raw trends for the bulge-dominated and disc-dominated samples, respectively. The fitted \(\hat{\mathbf e}_{\rm sys}\) component captures the observed dependence of both \(\langle e_1\rangle\) and \(\langle e_2\rangle\) on PSF ellipticity, FWHM, airmass, limiting magnitude, PSF size, Galactic reddening and sky brightness variance. Extended Data Figures~\ref{fig:bulge_e_sys_residual} and \ref{fig:disc_e_sys_residual} show the same diagnostics after subtracting \(\hat{\mathbf e}_{\rm sys}\). The residual trends are consistent with the residual-null expectation across all PSF and survey-condition quantities considered. The reduced \(\chi^2/\nu\) values shown in the panels quantify this comparison between the residual data and the residual-null ensemble.

This diagnostic also addresses the possibility that the \(\sigma_\theta\) selection could bias the position-angle distribution through PSF or survey-condition structure. The test is performed on the final selected catalogues, after all cuts. Therefore, any preferential selection of galaxies whose measured position angles align with the PSF or with observing-condition gradients would appear as a non-zero residual trend of \(\langle e_1\rangle\) or \(\langle e_2\rangle\) with the corresponding diagnostic. The absence of such residual trends shows that PSF- and survey-condition-induced alignments are either not present at the relevant level or are explicitly modelled and subtracted before the LAIA estimator is applied.

Possible coupling to the local density field is assessed by tests with $N$-body simulations, redshift-split measurements, use of different large-scale-structure tracers, and comparisons across widely separated sky regions. These tests show the success of our estimator (see main text), such that the measured signal persists under changes in redshift range, tracer definition and sky footprint.

We also use the residualised catalogues to construct DES-based empirical null catalogues. For each galaxy sample, we generate \(N_{\rm null}=500\) independent null realizations. These nulls are designed to preserve the survey footprint, redshift distribution, \(\sigma_\theta\) distribution, galaxy selection function, PSF and survey-condition values, and the object-by-object residual ellipticity amplitude. For each galaxy we define the residual amplitude
\begin{equation}
    R_i
    =
    \left[
        e_{{\rm res},1,i}^2
        +
        e_{{\rm res},2,i}^2
    \right]^{1/2}.
\end{equation}
For each null realization \(m=1,\ldots,N_{\rm null}\), we draw an independent headless angle
\begin{equation}
    \phi_i^{(m)}\sim{\cal U}(0,\pi),
\end{equation}
and define a randomized residual vector
\begin{equation}
    r_{1,i}^{(m)}
    =
    R_i\cos 2\phi_i^{(m)},
    \qquad
    r_{2,i}^{(m)}
    =
    R_i\sin 2\phi_i^{(m)} .
    \label{eq:randomized_residual}
\end{equation}
The corresponding full empirical null ellipticity is
\begin{equation}
    e_{{\rm fullnull},j,i}^{(m)}
    =
    \hat e_{{\rm sys},j,i}
    +
    r_{j,i}^{(m)},
    \qquad j=1,2 .
    \label{eq:full_empirical_null}
\end{equation}
This form preserves the fitted PSF and survey-condition component while randomizing the residual orientation. For analyses performed on the residualised catalogues, we subtract the same fitted systematic component and obtain the residual empirical null
\begin{equation}
    e_{{\rm resnull},j,i}^{(m)}
    \equiv
    r_{j,i}^{(m)} .
    \label{eq:residual_empirical_null}
\end{equation}
Thus the residual nulls reproduce the residual amplitude distribution and all catalogue-level selection properties of the data, but contain no coherent axial-alignment signal by construction.

As an external validation of this empirical null construction, we compare the resulting \(\theta_{\rm IA}\) distribution with the corresponding null distributions obtained from the Euclid Flagship 2 and MICECAT v2.0 $N$-body-based mock catalogues in Fig.~\ref{fig:results}. The DES-based residual empirical nulls reproduce the same broad \(\theta_{\rm IA}\) behaviour seen in those cosmological mocks, indicating that the residual randomisation captures the relevant orientation-noise budget of the selected samples. Because the Euclid Flagship 2 nulls yield slightly more conservative significances, the headline significance values reported in the main text are based on the Euclid Flagship 2 mock ensemble. The DES-based empirical residual nulls are used for controlled LAIA injection tests; see the LAIA injection pipeline in Methods~\ref{sec:naive_mocks}. They are also used in the residual-bias tests in Methods~\ref{sec:residualbias} and in the coherence-scale test in Methods~\ref{sec:coherence}.

\subsection{Permutation nulls for filament orientations}
\label{app:filament_nulls}

We assess the significance of the filament measurements with empirical permutation null catalogues generated on the observed filament footprint. The nulls are constructed for the full segment samples used in the main analysis: FN DR16, FS DR16 and FN DR12. For each sample, we generate 500 null realizations and apply the same LAIA estimator used for the observed segment catalogue.

The permutation is performed at the level of parent filaments rather than individual segments. For each parent filament, we compute a mean axial orientation from its segment tangent vectors after transporting them to a common tangent plane at the filament centre. Parent filaments are then assigned to local permutation pools defined by a HEALPix sky patch and by the redshift slice encoded in the filament catalogue. In both catalogues, the filament reconstruction is provided in discrete redshift slices; in our segment tables these slices are labelled by the catalogue redshift interval, \((z_{\rm low},z_{\rm high})\), and we retain these native catalogue slices when defining the permutation pools, rather than introducing new redshift bins. The sky patches are defined with HEALPix NSIDE \(=4\), corresponding to \(214.9\,{\rm deg}^2\) per patch, or an effective angular scale of about \(14.7^\circ\).

Within each sky-patch--redshift-slice pool, the mean axial orientations are randomly permuted between parent filaments. The donor orientation is transported to the recipient filament and applied as a common axial rotation to all of its segments. This preserves the footprint, full-sample redshift distribution, native redshift-slice structure, number of segments per parent filament, local one-point orientation distribution and internal segment-to-segment coherence, while removing the large-scale association between sky position and absolute filament orientation. 

\subsection{\label{sec:psf}Estimator-level PSF leakage and empirical-null base-level correction}

The residualisation procedure described in Section~\ref{sec:residual_correlations} acts at the catalogue level: it removes, galaxy by galaxy, the component of the observed spin-2 orientation field that is spuriously correlated with PSF and survey-condition diagnostics. As a final safeguard, we also perform an estimator-level correction. This step does not modify individual galaxy orientations. Instead, it is applied directly to the LAIA estimator maps \(E^{\rm X}(\hat{\mathbf d})\), after the residualised galaxy catalogue has been passed through the estimator.

This correction is defined only for the galaxy-orientation part of the measurement. The PSF template is constructed from the PSF semi-axes at the positions of the galaxies and is used to test whether the galaxy-based LAIA estimator retains any residual projection along the PSF pattern. The filament or large-scale-structure tracer catalogues are not corrected against a PSF template; they enter the analysis only through the geometric directions used by the estimator.

Let \(E_{\rm obs}^{\rm X}(\hat{\mathbf d})\) denote the LAIA estimator map computed from the residualised galaxy catalogue, where \(\hat{\mathbf d}\) runs over the tested directions, taken to be the centres of \texttt{HEALPix} pixels at \(\texttt{NSIDE}=32\). The index \({\rm X}\in\{BD,DD\}\) labels the two estimator projections defined in Eq.~\eqref{eq:estimator}. For each galaxy sample we also compute the same estimator on the \(N_{\rm null}=500\) empirical residual-null catalogues described in Section~\ref{sec:residual_correlations}. This gives a set of null maps \(E_{{\rm null},m}^{\rm X}(\hat{\mathbf d})\), with \(m=1,\ldots,N_{\rm null}\).

We first subtract the mean null response of the estimator,
\begin{equation}
    \bar E_{\rm null}^{\rm X}(\hat{\mathbf d})
    =
    \frac{1}{N_{\rm null}}
    \sum_{m=1}^{N_{\rm null}}
    E_{{\rm null},m}^{\rm X}(\hat{\mathbf d}) .
    \label{eq:null_base_level}
\end{equation}
This quantity is the estimator base level expected in the absence of a coherent axial-alignment signal, given the same survey footprint, galaxy positions, redshift distribution, residual ellipticity amplitudes, \(\sigma_\theta\) distribution and selection function as the data. In an ideal isotropic and infinitely sampled catalogue this map would vanish after averaging over null realizations. In a finite survey footprint, however, the empirical nulls can retain a small anisotropic estimator response. We verified that this mean null map is statistically negligible compared with both the observed signal and the null dispersion, but subtract it for completeness:
\begin{equation}
    E_{0}^{\rm X}(\hat{\mathbf d})
    =
    E_{\rm obs}^{\rm X}(\hat{\mathbf d})
    -
    \bar E_{\rm null}^{\rm X}(\hat{\mathbf d}) .
    \label{eq:base_level_corrected}
\end{equation}
Thus \(E_{0}^{\rm X}\) is the observed estimator map after removal of the empirical-null base level.

We then test for any remaining large-scale PSF leakage at the estimator level. For this purpose, we construct a PSF-template estimator map, \(E_{\rm PSF}^{\rm X}(\hat{\mathbf d})\), by applying the same LAIA estimator to the PSF semi-axes at the positions of the galaxies. This PSF map is not used as a physical alignment field; it is a template for the response that would be produced if the estimator were sensitive to residual PSF anisotropy.

All maps entering this projection are mean-subtracted over the tested directions,
\begin{equation}
    \tilde E^{\rm X}(\hat{\mathbf d})
    =
    E^{\rm X}(\hat{\mathbf d})
    -
    \left\langle E^{\rm X}\right\rangle_{\hat{\mathbf d}},
\end{equation}
where \(\langle\cdot\rangle_{\hat{\mathbf d}}\) denotes an average over the \(\texttt{NSIDE}=32\) direction grid. We model the base-level-corrected estimator map as
\begin{equation}
    \tilde E_{0}^{\rm X}(\hat{\mathbf d})
    =
    \tilde E_{\rm IA}^{\rm X}(\hat{\mathbf d})
    +
    \eta_{R}^{\rm X}
    \tilde E_{\rm PSF}^{\rm X}(\hat{\mathbf d}) .
    \label{eq:psf_model_real}
\end{equation}
Here \(\tilde E_{\rm IA}^{\rm X}\) is the PSF-deleaked estimator map, and \(\eta_{R}^{\rm X}\) is the residual leakage coefficient. This model removes only the component of the estimator map that is colinear with the PSF template. Since the true astrophysical alignment field is expected to be statistically uncorrelated with the instrumental PSF pattern, this projection is unbiased in expectation.

The leakage coefficient is obtained by a one-parameter least-squares fit,
\begin{equation}
    \hat\eta_{R}^{\rm X}
    =
    \arg\min_{\eta}
    \left\langle
    \left[
        \tilde E_{0}^{\rm X}(\hat{\mathbf d})
        -
        \eta\,
        \tilde E_{\rm PSF}^{\rm X}(\hat{\mathbf d})
    \right]^2
    \right\rangle_{\hat{\mathbf d}} .
    \label{eq:eta_ls_real}
\end{equation}
Equivalently, when
\(\mathrm{Var}(\tilde E_{\rm PSF}^{\rm X})>0\),
\begin{equation}
    \hat\eta_{R}^{\rm X}
    =
    \frac{
    \mathrm{Cov}
    \left(
        \tilde E_{0}^{\rm X},
        \tilde E_{\rm PSF}^{\rm X}
    \right)}
    {
    \mathrm{Var}
    \left(
        \tilde E_{\rm PSF}^{\rm X}
    \right)
    } .
    \label{eq:eta_cov_real}
\end{equation}
The final estimator-level corrected map is therefore
\begin{equation}
    \tilde E_{\rm corr}^{\rm X}(\hat{\mathbf d})
    =
    \tilde E_{0}^{\rm X}(\hat{\mathbf d})
    -
    \hat\eta_{R}^{\rm X}
    \tilde E_{\rm PSF}^{\rm X}(\hat{\mathbf d}) .
    \label{eq:psf_deleaked_map}
\end{equation}
The preferred LAIA direction, \(\hat{\mathbf d}_{\rm IA}\), and the corresponding amplitude, \(\theta_{\rm IA}\), are then computed from the pair of corrected maps
\(\tilde E_{\rm corr}^{BD}\) and \(\tilde E_{\rm corr}^{DD}\).

The two estimator-level subtractions have different roles. The empirical-null base-level correction removes the mean response of the estimator expected from the survey mask, finite sampling, selection function and residual amplitude distribution in the absence of coherent alignment. The PSF projection then removes any remaining component of the base-level-corrected estimator map that is aligned with the PSF template. The fitted residual PSF leakage coefficients are small, with amplitudes of order \(\hat\eta_{R}^{\rm X}\simeq0.02\), but we apply the correction uniformly as a conservative final de-leakage step. This ensures that the reported LAIA measurements are based on residualised galaxy orientations and on estimator maps that have also been orthogonalized against the PSF response and corrected for the empirical-null base level.

\subsection{\label{sec:naive_mocks}LAIA injection and naive mock generation}

We verify the estimator and perform controlled systematics tests by injecting an axial LAIA signal of known amplitude $\theta_{\rm IA}$ about a fixed sky direction $\hat{\mathbf d}_{\rm IA}$. This injection procedure is used in two ways. First, we generate naive control mocks, in which random position angles are placed on the observed DES angular mask and galaxy-density field but without the full DES morphology and quality-selection structure. These mocks test whether the footprint and estimator alone can generate residual biases. Second, we inject LAIA into the DES-based empirical null catalogues used in the coherence-scale test (Methods~\ref{sec:coherence}), the residual-bias calibration (Methods~\ref{sec:residualbias}) and the morphology-misclassification tests (Methods~\ref{sec:misclassification}). These tests are diagnostic rather than cosmological: they measure the response of the estimator and analysis pipeline to a known injected signal.

For each object, we draw an intrinsic position angle $\theta_i \sim \mathcal{U}(0,\pi)$, constructing the orthonormal tangent-plane as in Eqs.~\ref{eq:semiaxis}, \ref{eq:spineaxis} and \ref{eq:uaxis}. When injecting into DES-based empirical null catalogues, the initial orientation is instead taken from the corresponding null realization, which preserves the residual ellipticity amplitudes, footprint, weights and catalogue selection properties. For every sky position, we then project the LAIA axis onto the local tangent plane,
\begin{equation}
    \hat{\mathbf d}^{\,\rm proj}_i \equiv
    \frac{\hat{\mathbf d}_{\rm IA} - (\hat{\mathbf d}_{\rm IA}\!\cdot\!\hat{\mathbf n}_i)\,\hat{\mathbf n}_i}
         {\big\lVert \hat{\mathbf d}_{\rm IA} - (\hat{\mathbf d}_{\rm IA}\!\cdot\!\hat{\mathbf n}_i)\,\hat{\mathbf n}_i \big\rVert}\,,
\end{equation}
and, prior to any rotation, define the initial axial separation
\begin{equation}
    \phi^{\rm orig}_i \equiv \arccos\!\left(\left|\hat{\mathbf u}_i^X \!\cdot\! \hat{\mathbf d}^{\,\rm proj}_i\right|\right),
\end{equation}
where the absolute value enforces the $180^\circ$ degeneracy appropriate for an axial major axis and $X=\{BD,DD,F\}$. The rotation amplitude is limited by the saturation prescription in Eq.~\ref{eq:saturation},
\begin{equation}
    \tilde{\phi}_i \equiv \min\!\left(\phi^{\rm orig}_i,\,\theta_{\rm IA}\right),
\end{equation}
so that an object closer to the alignment axis than $\theta_{\rm IA}$ is not rotated through it. To choose the shortest axial rotation, the projected LAIA direction is replaced by its antipode whenever necessary so that it lies in the same axial hemisphere as the tracer orientation. The sense of the rotation on the tangent plane is then fixed by the scalar triple product,
\begin{equation}
    s_i \equiv \operatorname{sgn}\!\big[\hat{\mathbf n}_i \!\cdot\! (\hat{\mathbf u}_i^X \times \hat{\mathbf d}^{\,\rm proj}_i)\big] \in \{-1,+1\},
    \quad
    \phi^{\prime}_i \equiv s_i\,\tilde{\phi}_i \, .
\end{equation}

We then rotate $\hat{\mathbf u}_i^X$ about $\hat{\mathbf n}_i$ by $\phi^{\prime}_i$ using Rodrigues' formula \cite{Rodrigues1840}. Because $\hat{\mathbf u}_i^X\!\cdot\!\hat{\mathbf n}_i=0$, the expression reduces to
\begin{equation}
    \hat{\mathbf u}^{X,\rm rot}_i
    = \hat{\mathbf u}_i^X \cos \phi^{\prime}_i
    + \left(\hat{\mathbf n}_i \times \hat{\mathbf u}_i^X\right)\sin \phi^{\prime}_i \, .
\end{equation}
For empirical-null injections that require ellipticity components rather than only PAs, we preserve the original residual ellipticity amplitude and replace only the headless orientation angle by the injected one. The lensing leakage is included as an extra stochastic axial rotation with fixed amplitude calibrated in Methods~\ref{sec:lensing}. 

For the naive mocks, measurement noise is included as an additional axial rotation,
\begin{equation}
    \phi^{\rm noise}_i \sim \mathcal N\!\left(0,\,\sigma_\theta^2(\hat{\mathbf n}_i)\right),
\end{equation}
where $\sigma_\theta(\hat{\mathbf n}_i)$ is the mean PA uncertainty of the sample in the sky pixel at \texttt{NSIDE}=128 containing $\hat{\mathbf n}_i$. For filament-spine injections, no PA-error attenuation is applied. 

These controlled injection procedures complement, but does not replace, the cosmological mock null tests. Euclid Flagship 2 and MICECAT v2.0 already test whether standard local IA prescriptions, sampled in statistically isotropic $\Lambda$CDM light cones and matched to the DES selection and PA-error distributions, can generate a footprint-scale LAIA signal. The injection tests instead ask a more targeted question: how the estimator responds to a known axial signal, what level of coherent angular structure is required for local patches to leak into a sky-scale measurement, and how residual biases or morphology misclassification propagate into the recovered LAIA direction and amplitude.

\subsection{\label{sec:coherence}Coherence test}

A possible concern is that the LAIA estimator could respond not only to a genuinely sky-scale orientation field, but also to the superposition of many local IA patches within a finite light cone. Our primary null tests already address this effect with Euclid Flagship 2 and MICECAT v2.0 mock catalogues. These mocks are built in statistically isotropic $\Lambda$CDM cosmologies and include standard local IA prescriptions; after resampling them to match the DES selection and PA-error distributions, they do not reproduce the observed LAIA amplitudes. Thus, standard local IA averaged over an isotropic light cone is not sufficient to explain the measured signal.

\begin{figure}[t]
    \includegraphics[width=0.99\linewidth]{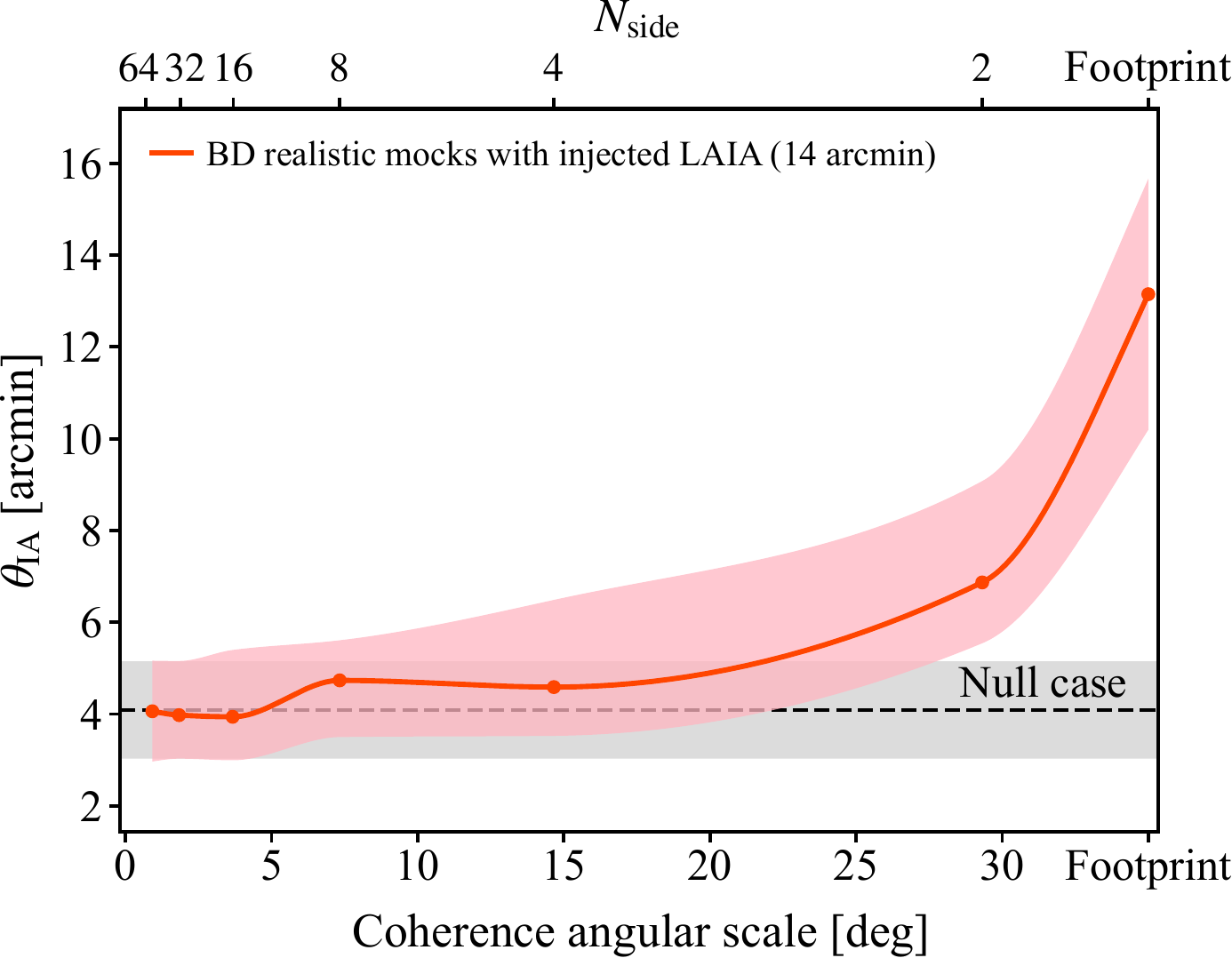}
\caption{
\textbf{Coherence-scale test.}
A controlled LAIA signal with $\theta_{\rm IA}=14'$ is injected into bulge-dominated DES-based empirical null catalogues. The injected alignment $\theta_{\rm IA}$ is coherent only within \texttt{HEALPix} blocks, with the block size set by the \texttt{NSIDE} value shown on the upper axis; each block is assigned an independent LAIA direction. Shaded regions represents $1\sigma$ error bars. The full-footprint point corresponds to a single coherent injected direction across the survey. Applying the estimator to the full footprint shows that local IA patches average out unless the injected field remains coherent over angular scales of order~$\gtrsim30^\circ$.
}
    \label{fig:coherence}
\end{figure}

As an additional controlled diagnostic, we perform a coherence-scale test. The purpose is not to model a physical IA field, but to determine the angular scale over which an injected alignment must remain coherent before it contributes appreciably to the footprint-scale LAIA estimator. We start from the bulge-dominated DES-based empirical null catalogues constructed from residual ellipticity components after removal of PSF and survey-condition correlations. We then inject a LAIA signal with $\theta_{\rm IA}=14'$, imposing coherence only within finite angular blocks. The recovered amplitudes as a function of the injected coherence scale are shown in Extended Data Figure~\ref{fig:coherence}.

The survey footprint is partitioned into \texttt{HEALPix} pixels with different \texttt{NSIDE} values. Each block is assigned an independent LAIA direction, and all galaxies within that block receive the corresponding coherent injected alignment (Methods~\ref{sec:naive_mocks}). We then apply the standard LAIA estimator to the full footprint, without using the block information. The limiting case labelled as the full footprint corresponds to a single coherent injected direction across the entire survey, whereas the other cases represent smaller local IA patches with unrelated directions.

The recovered amplitudes remain consistent with the null expectation for small coherence patches. A significant LAIA response appears only when the injected orientation field is coherent over angular scales of order $\gtrsim30^\circ$ (Extended Data Figure~\ref{fig:coherence}). This demonstrates that the estimator is not substantially affected by a simple superposition of small-scale or moderately local IA regions with unrelated axes. Instead, producing a footprint-scale LAIA signal requires coherence over very large angular scales. This result complements the $N$-body mock comparison by isolating the coherence-scale requirement in a minimal, controlled setting.
\vspace{0.5cm}

\subsection{\label{sec:residualbias}Residual bias}

After subtracting the ellipticity components correlated with PSF quantities and survey-condition maps, we perform a final injection-based test to determine whether the LAIA pipeline leaves any residual bias in the recovered direction or amplitude. This test is designed to capture biases that are not visible as simple one-point correlations with survey diagnostics, but that can arise from the combination of the DES footprint, galaxy selection function, position-angle uncertainties and the nonlinear maximization of the LAIA estimator.

\begin{figure*}[t]
    \includegraphics[width=0.89\textwidth]{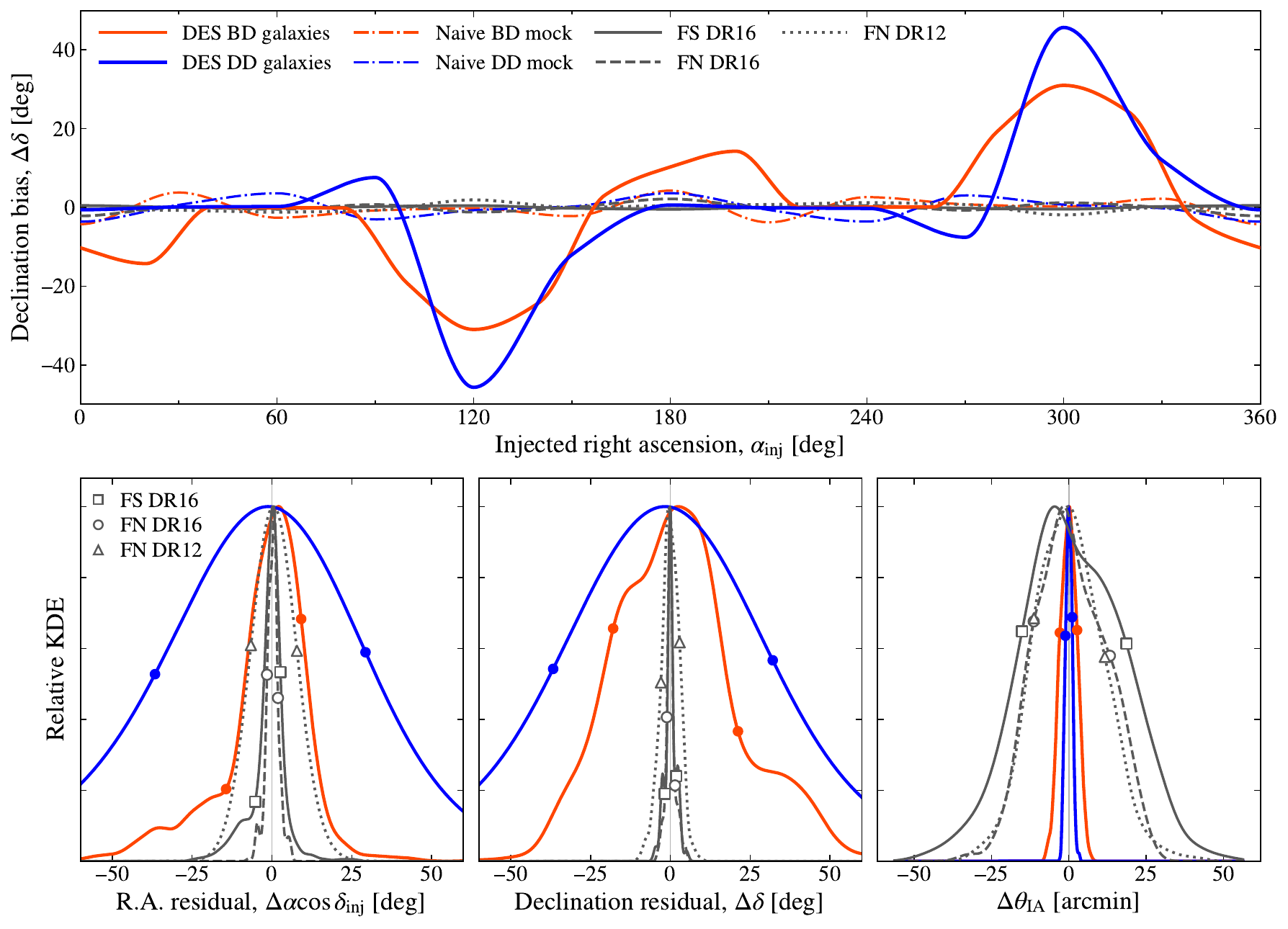}
    \caption{
    \textbf{Residual-bias calibration.}
    LAIA signals are injected into empirical null catalogues and recovered with the full estimator pipeline to test for residual biases in $\alpha$, $\delta$ and $\theta_{\rm IA}$. Galaxy tests use 3000 injected realizations with isotropically distributed input axes; filament tests use 1000 injected realizations.
    \textbf{Top:} residual declination bias for the galaxy measurements as a function of right ascension. This RA-dependent bias is subtracted from the reported galaxy LAIA directions.
    \textbf{Bottom left:} residual right-ascension bias, consistent with zero.
    \textbf{Bottom centre:} recovered declination residuals after the RA-dependent declination-bias correction. The mean residual is consistent with zero, although the scatter remains direction-dependent because different sky directions are constrained differently by the DES footprint and selection function.
    \textbf{Bottom right:} residual bias in the recovered LAIA amplitude. No significant estimator-induced $\theta_{\rm IA}$ bias is detected, apart from the small bulge-dominated amplitude shift associated with morphology misclassification, which is corrected separately in Methods~\ref{sec:misclassification}. Filament measurements show no residual bias in $\alpha$, $\delta$ or $\theta_{\rm IA}$.
    }
    \label{fig:residual_bias}
\end{figure*}

We use the LAIA injection procedure described in Methods~\ref{sec:naive_mocks}. For the galaxy samples, we inject LAIA signals into DES-based empirical null catalogues constructed from the residual ellipticity components after removal of the PSF- and survey-condition-correlated component. We use 3000 injected realizations with input LAIA axes distributed isotropically over the sky. For the filament samples, we perform an analogous validation with 1000 injected realizations. In each case, we compare the injected and recovered LAIA parameters,
\begin{align}
    \Delta\alpha &=
    \alpha_{\rm rec}-\alpha_{\rm inj}, \nonumber
    \\
    \Delta\delta &=
    \delta_{\rm rec}-\delta_{\rm inj},
    \\
    \Delta\theta_{\rm IA} &=
    \theta_{{\rm IA},\,{\rm rec}}-\theta_{{\rm IA},\,{\rm inj}}, \nonumber
\end{align}
after applying the same axial convention used for the data. Extended Data Figure~\ref{fig:residual_bias} summarizes the resulting residual biases. We find no significant residual bias in right ascension and $\theta_{\rm IA}$ for any tracer. 

For filaments, the same injection validation shows no significant residual bias in right ascension, declination or $\theta_{\rm IA}$. We therefore apply no residual-bias correction to the filament measurements. This is consistent with the fact that filament orientations are not image-level shape measurements and are not affected by PSF, ellipticity or morphology-selection effects.

For the galaxy measurements, however, we detect a residual bias in the recovered declination that depends on right ascension. This bias remains after subtraction of the fitted PSF and survey-condition component, and is therefore treated as a final estimator-level calibration. We measure the mean declination residual as a function of right ascension from the injected catalogues and subtract this calibration from the galaxy LAIA directions reported in the main text.

The origin of this residual declination bias is associated with the DES selection, systematics and measurement structure rather than with the LAIA estimator alone. To verify this, we repeat the same injection test on naive control mocks that preserve the DES angular mask and galaxy-density field but do not include the full DES realistic residual ellipticity components. These naive mocks do not show the same residual declination bias. This indicates that the effect is not induced by the footprint geometry alone.

This residual bias is not a candidate explanation for any fraction of the LAIA signal itself. It affects only the recovered declination of galaxy-based measurements, is corrected in the reported galaxy axes, and is not seen in right ascension or in the recovered $\theta_{\rm IA}$ amplitude. It is also absent in the filament measurements, which are based on independent tracers and a different survey footprint. Moreover, such a one-dimensional residual cannot reproduce the morphology-dependent BD--DD axial pattern, the redshift evolution of the bulge-dominated axis, or the agreement with independent filament catalogues. We therefore treat it as a calibration-level correction to the recovered galaxy direction. Unlike ground-based surveys, space telescopes such as CSST and \textit{Euclid} will not suffer from the same atmosphere-driven declination-dependent bias, as indicated by the naive mocks.

After applying the declination-bias correction, the mean residual in recovered declination is consistent with zero. The scatter in recovered declination, however, remains direction-dependent: some injected sky directions are intrinsically less well constrained by the DES footprint and selection than others. This anisotropic recovery variance is propagated into the reported galaxy uncertainties through the spatial block bootstrap and the injection-based calibration. By contrast, the recovered right ascension and $\theta_{\rm IA}$ show no comparable direction-dependent residual bias. The only relevant amplitude bias is the small bulge-dominated $\theta_{\rm IA}$ shift associated with morphology misclassification, which is treated separately in Methods~\ref{sec:misclassification} and included in the final reported measurements.
\newpage
\subsection{\label{sec:misclassification} Misclassification bias}

The contamination fraction $\kappa$ was obtained as the average of the probability of the galaxies being late-type (disc-dominated) or early-type (bulge-dominated) from two different algorithms, we name the probabilities $P^{\rm V}$ obtained by \citet{DES:2020tkt} and $P^{\rm C}$ by \citet{2021MNRAS.507.4425C}:
\begin{equation}
    \kappa^{\rm Y}=1-\sum_{i=1}^{N^{\rm Y}_{\rm gal}} \frac{P^{\rm V, Y}_i + P^{\rm C,Y}_i}{2N^{\rm Y}_{\rm gal}}\, , 
\end{equation} 
where ${\rm Y}=\{l,e\}$, for late-type ($l$) and early-type ($e$).
We estimate 0.27\% of contamination for disk-dominated galaxies (in DD). The BD contamination is much bigger, 26.55\%. In order to understand this difference, it is important to notice that both \cite{DES:2020tkt} and \cite{2021MNRAS.507.4425C} deep learning methods are trained using Galaxy Zoo visual morphology, which relies on the ability of humans to identify patterns on images. Spiral structures are readily recognized through visual inspection up to a given threshold, for example missing faint spiral structures. Moreover, lenticular galaxies, which are characterized by a disk and a prominent bulge, can be associated to both early and late type class. In few words, while the late type class encompass all objects with visible spiral arms, the early type class might contain elliptical galaxies, lenticular galaxies, multiple object images, low surface brightness galaxies,
bulge-dominated spiral galaxies, or faint/compact spiral galaxies, in different numbers in the two papers, resulting in a higher number of discordant classification and, therefore, a bigger associated error \citep{Lintott:2008ne,2024MNRAS.528.4188B}.

The impact of this contamination is negligible for the DD sample, for which the inferred contamination fraction is well below the statistical sensitivity of our analysis; we therefore do not apply any correction to the DD measurements. For the BD sample, the larger contamination fraction requires an explicit calibration. We quantified its effect through injection tests in which a LAIA signal was added to null realizations, while including the expected fraction of contaminants as galaxies carrying no LAIA contribution. This procedure measures the dilution response of the estimator to the inferred BD contamination. We find that the contamination reduces the recovered value of $\theta_{\rm IA}$ by $1.8'$--$2.5'$, depending on the redshift cut adopted. The BD values of $\theta_{\rm IA}$ reported in the main text include this contamination correction.

\subsection{\label{sec:error} Spatial block bootstrap}

We estimate uncertainties with a spatial block bootstrap at the \texttt{HEALPix} pixel level. We partition the sky into equal-area pixels ($\texttt{NSIDE}=8$), identify the set of occupied pixels within the survey footprint, and treat each pixel as one resampling unit. For each of realizations we sample pixels with replacement, drawing the same number of pixel as in the data from the set of occupied pixels (each pixel equally likely). For every selected pixel, we include all galaxies/filament segments contained in that pixel (with their per-object uncertainties and matched PSF entry). Resampling whole pixels preserves local clustering and position-dependent selection/PSF systematics at scales bigger than the pixel size ($\sim7.3^{\circ}$), providing a conservative uncertainty that respects survey inhomogeneities, something an object-level bootstrap would not capture as well. To test the boostrap procedure we rely on the mocks generated as discussed in Methods~\ref{sec:coherence},\ref{sec:naive_mocks} and \ref{sec:residualbias}, verifying that they recover the expected value on average and similar variance.

\subsection{\label{app:bulkflow_laia_relation}Bulk-flow directions and the low-redshift LAIA axis}

The low-redshift LAIA direction and bulk-flow axes are sourced by different derivatives of the same large-scale gravitational potential: the peculiar velocity probes the gradient of the potential, whereas intrinsic alignments respond to its trace-free Hessian. These derivative operators do not select the same direction for a generic potential. A correlation is expected only in the more restricted case in which the low-redshift gravitational field is dominated by a single large-scale axis, such as an approximately colinear attractor--repeller configuration.

Let \(\Phi(\mathbf{x},a)\) be the peculiar Newtonian gravitational potential at comoving position \(\mathbf{x}\) and scale factor \(a\). In linear theory, the growing-mode peculiar velocity \(\mathbf{v}\) is proportional to the gravitational acceleration \(\mathbf{g}\),
\begin{equation}
    \mathbf{v}(\mathbf{x},a)
    \propto
    \mathbf{g}(\mathbf{x},a),
    \qquad
    g_i(\mathbf{x},a)
    \equiv
    -\partial_i\Phi(\mathbf{x},a).
    \label{eq:v_grad_phi}
\end{equation}
The proportionality factor is positive for the growing mode and is irrelevant here, because we only compare directions. The tidal field relevant for intrinsic alignments is the trace-free part of the Hessian of the same potential. We define the tidal-stretching tensor as
\begin{equation}
    {\cal T}_{ij}(\mathbf{x},a)
    \equiv
    -\left[
    \partial_i\partial_j\Phi(\mathbf{x},a)
    -
    \frac{1}{3}\delta_{ij}\nabla^2\Phi(\mathbf{x},a)
    \right],
    \label{eq:tidal_tensor_def}
\end{equation}
up to an arbitrary positive normalization. The overall sign is a convention; the comparison below uses only the principal headless axis of \({\cal T}_{ij}\), since our LAIA estimator identifies \(\hat{\mathbf d}\equiv-\hat{\mathbf d}\). In the tidal-alignment interpretation, this axis is the projected coherent tidal direction traced statistically by galaxy position angles \citep{CatelanKamionkowskiBlandford2001,Joachimi:2015mma,Kiessling:2015sma,Troxel:2014dba}.

The observational motivation for considering an attractor--repeller geometry comes from reconstructions of the local velocity field. Cosmicflows analyses indicate that the nearby flow is strongly shaped by an attractor associated with the Shapley concentration and by large underdense regions acting as effective repellers, including the Dipole Repeller \citep{Tully:2014Nature,Hoffman:2017NatAs,Courtois:2017ApJL,Pomarede:2017ApJ}. In particular, \citet{Hoffman:2017NatAs} found that the bulk-flow dipole is closely anti-aligned with the Dipole Repeller, while the expansion eigenvector of the velocity-shear quadrupole is closely aligned with the Shapley Attractor. This motivates the idealized calculation below: a distant, approximately spherical attractor or repeller.

We choose the origin \(\mathbf{x}=0\) to be the point around which the local expansion is made, for example the observer in the bulk-flow comparison. This choice is only a convention: the same calculation holds around any other point after redefining \(\mathbf{x}\) as the displacement from that point. Let the dominant structure be centred at
\begin{equation}
    \mathbf{R}=R\hat{\mathbf d},
\end{equation}
where \(R\) is its distance from the origin and \(\hat{\mathbf d}\) is the corresponding unit direction. If the structure is approximately spherical after averaging over its internal substructure, its contribution to the potential can be written as
\begin{equation}
    \Phi(\mathbf{x})
    =
    \psi(|\mathbf{x}-R\hat{\mathbf d}|),
    \label{eq:spherical_source_potential}
\end{equation}
where \(\psi(r)\) is an arbitrary radial potential profile. This is the monopole approximation to the potential of a dominant attractor or repeller. Departures from spherical symmetry, additional structures and the survey window produce scatter around the idealized result.

For \(|\mathbf{x}|\ll R\), evaluated at the origin, the gradient is
\begin{equation}
    \left.\partial_i\Phi\right|_{\mathbf{x}=0}
    =
    -\psi'(R)\hat d_i.
    \label{eq:source_gradient}
\end{equation}
The gravitational acceleration is therefore
\begin{equation}
    g_i(0)
    =
    -\left.\partial_i\Phi\right|_{\mathbf{x}=0}
    =
    \psi'(R)\hat d_i,
    \label{eq:source_acceleration}
\end{equation}
so the induced bulk-flow contribution is parallel or antiparallel to \(\hat{\mathbf d}\), depending on the sign of \(\psi'(R)\), i.e. on whether the source acts as an attractor or as an effective repeller.

The Hessian of the same potential is
\begin{equation}
    \left.\partial_i\partial_j\Phi\right|_{\mathbf{x}=0}
    =
    \psi''(R)\hat d_i\hat d_j
    +
    \frac{\psi'(R)}{R}
    \left(
    \delta_{ij}-\hat d_i\hat d_j
    \right).
    \label{eq:source_hessian}
\end{equation}
Using Eq.~(\ref{eq:tidal_tensor_def}), its trace-free tidal part becomes
\begin{equation}
    \!{\cal T}_{ij}(0)
    =
    A_R
    \left(\!
    \hat d_i\hat d_j\!-\!\frac{1}{3}\delta_{ij}
    \!\right),
    \;\;
    A_R
    \equiv
    -\left[\!
    \psi''(R)\!-\!\frac{\psi'(R)}{R}
    \!\right].
    \label{eq:source_tidal_axis}
\end{equation}
This equation shows explicitly that \(\hat{\mathbf d}\) is an eigenvector of the tidal tensor. Contracting Eq.~(\ref{eq:source_tidal_axis}) with \(\hat d_j\), we obtain
\begin{align}
    {\cal T}_{ij}\hat d_j
    &=
    A_R
    \left(
    \hat d_i\hat d_j\hat d_j
    -
    \frac{1}{3}\delta_{ij}\hat d_j
    \right) \nonumber \\
    &=
    A_R
    \left(
    \hat d_i
    -
    \frac{1}{3}\hat d_i
    \right)
    =
    \frac{2}{3}A_R\hat d_i ,
    \label{eq:tidal_eigen_parallel}
\end{align}
where we used the sum \(\hat d_j\hat d_j=1\). Since the result is proportional to \(\hat d_i\), the direction \(\hat{\mathbf d}\) is a principal tidal axis. It is also the unique non-degenerate axis selected by the source, with the transverse plane remaining degenerate. If \(A_R>0\), this axis is the stretching direction in the convention of Eq.~(\ref{eq:tidal_tensor_def}); if \(A_R<0\), it is the compression direction. In both cases, the headless tidal axis is the same. Since LAIA is axial and the IA response sign can depend on morphology, the relevant comparison is the axis, not the sign of the eigenvalue.

A colinear attractor--repeller pair is a linear superposition of such contributions along the same \(\hat{\mathbf d}\). Unless the tidal amplitudes cancel accidentally, the resulting tensor keeps the same form as Eq.~(\ref{eq:source_tidal_axis}), with a different value of \(A_R\). Therefore, in the single-dominant-axis limit motivated by local-flow reconstructions, the bulk-flow vector and the LAIA tidal axis are different derivative projections of the same large-scale gravitational configuration. The expected relation is directional correlation, not exact equality.

This expectation is most relevant for our lowest-redshift galaxy split. The bin \(z\leq0.25\) should not be interpreted as a purely local-volume measurement: many galaxies in this bin lie beyond the classical scale of individual nearby attractors. We are therefore not claiming that every galaxy at \(z\simeq0.2\) is dynamically affected by the present-day Shapley--Dipole-Repeller flow. The more conservative statement is that, relative to higher-redshift splits, the \(z\leq0.25\) sample gives greater statistical weight to nearby structures and long-wavelength modes that subtend large angles on the sky and are less diluted by line-of-sight averaging. If the same low-redshift, large-scale attractor--repeller axis contributes appreciably to both the local velocity dipole and the coherent tidal field sampled by galaxy orientations, then a correlation between the galaxy LAIA axis and local bulk-flow directions is physically plausible. In a realistic cosmic web, multiple attractors, repellers, nonlinear structures, projection effects and the survey window broaden this relation. The prediction is therefore statistical colinearity, not exact equality between the LAIA and bulk-flow axes.

Filament orientations are expected to be less directly susceptible to a coherent bulk flow than galaxy position angles in the lowest-redshift sample. A bulk flow is primarily a coherent motion of a region as a whole. Such a coherent translation changes the positions of the filament network but does not, by itself, rotate filament tangents or change the orientation of the cosmic-web skeleton. Reorienting a filament requires velocity gradients, tidal shear or nonlinear rearrangement of the density field, not merely a common motion of the structures embedded in it. This is consistent with the standard picture in which filaments are seeded by the initial density and tidal fields, with nonlinear evolution sharpening the web rather than randomly reorienting it \citep{BondKofmanPogosyan1996,Cautun:2014,Libeskind:2018}. Recent simulations further show that several comoving filament statistics, including length functions and overdensity profiles, evolve only weakly once the Hubble expansion is factored out \citep{GalarragaEspinosa:2024}. The stability of the filament LAIA direction across redshift is therefore natural: filaments trace a slowly evolving, integrated tidal skeleton, whereas the lowest-redshift galaxy sample LAIA measurement can be more sensitive to the same local large-scale modes that appear in bulk-flow measurements.
\vspace{1.4cm}


\bibliographystyle{apsrev4-2}
\bibliography{lss_ia}

\begin{acknowledgments}
{
We thank Dr.\ Chihway Chang for the many insightful discussions that greatly enriched this study, and Dr.\ Ting-Yun Cheng for generously granting us access to her galaxy morphology catalog. We are also grateful to Dr.\ Marco Gatti, Dr.\ Erin Sheldon and Dr.\ Bj\"{o}rn Malte Sch\"{a}fer for their helpful comments. We thank the anonymous referees for constructive comments and suggestions that improved the manuscript.
\textbf{Funding:} PDSF acknowledges the support from the start-up funding of Zhejiang University, Zhejiang provincial top level research support program, FAPES (Brazil) and FINEP/FACC (contract 01.22.0505.00). ROR thanks PIBIC/CNPq for financial support. PSF acknowledges the support from the start-up funding of Zhejiang University, Zhejiang provincial top level research support program, and Brazilian funding agency CNPq for financial support PCI-DB 316199/2025-7. VM acknowledges partial support from CNPq (Brazil) and FAPES (Brazil). AC aknowledges the FAPERJ (Brazil) grants E26/202.607/2022 and 210.371/2022(270993) and CNPq. CRB acknowledges the financial support from CNPq (316072/2021-4), FAPERJ (grants 201.456/2022 and 210.330/2022) and the FINEP/FACC (contract 01.22.0505.00). RC acknowledges the support from the start-up funding of Zhejiang University and Zhejiang provincial top level research support program.
\textbf{Scientific Illustration:} Fig.~\ref{fig:ia_illustration} was commissioned by the artist \href{https://odresilva.artstation.com/}{``Drê Silva''}.
\textbf{CosmoHub:} This work has made use of CosmoHub, developed by PIC (maintained by IFAE and CIEMAT) in collaboration with ICE-CSIC. It received funding from the Spanish government (grant EQC2021-007479-P funded by MCIN/AEI/10.13039/501100011033), the EU NextGeneration/PRTR (PRTR-C17.I1), and the Generalitat de Catalunya. \textbf{Astro Data Lab:} This research uses services or data provided by the Astro Data Lab, which is part of the Community Science and Data Center (CSDC) Program of NSF NOIRLab. NOIRLab is operated by the Association of Universities for Research in Astronomy (AURA), Inc. under a cooperative agreement with the U.S. National Science Foundation.
\textbf{SilkRiver cluster:} The authors would like to acknowledge the use of the SilkRiver Supercomputer of Zhejiang University (China).
\textbf{Sci-Com cluster:} The authors would like to acknowledge the use of the computational resources provided by the \href{https://computacaocientifica.ufes.br/scicom}{Sci-Com Lab} of the Department of Physics at UFES, which was funded by FAPES, CAPES, and CNPq. \textbf{Milliways HPC IF-UFRJ:} This work made use of the Milliways HPC computer located at the Instituto de Física in the Universidade Federal do Rio de Janeiro, managed and funded by \href{https://sites.google.com/view/arcos-ufrj/about?authuser=0}{ARCOS} (Astrophysics, Relativity and COSmology research group). \textbf{LRG HPC:}
This work made use of the LRG HPC computer, which was funded by FAPERJ (E-26/210.371/2022).  \textbf{CHE/Milliways cluster:} We acknowledge the use of the computational resources of the joint CHE/Milliways cluster, supported by a FAPERJ grant E-26/210.130/2023.
\textbf{DES Shape Catalog:} This work made use of the DES shape catalogue \cite{DES:2020ekd} (Year 3 Gold). \textbf{DES Morphology Catalogs:} This work made use of the DES morphology catalogs \cite{DES:2020tkt} and \cite{2021MNRAS.507.4425C} (Year 3 Gold). \textbf{SDSS Filament Catalogs:} This work made use of the SDSS filament catalogs \cite{Duque:2021xgw} (DR 17) and \cite{Chen:2015bra} (DR 12). \textbf{$N$-body mock catalogs:} This work made use of the Euclid Flagship 2 \cite{Euclid:2024few,Euclid:2026bur,Potter:2016ttn,2016ascl.soft09016P,Behroozi:2011ju} (with IA incluced), MICECAT v2 \cite{DES:2022aeh,mice1,mice2,mice3,mice4,mice5} (with IA incluced), and BUZZARD 2 \cite{buzzard2,buzzard1} mocks. 
\textbf{DES:} This project used public archival data from the Dark Energy Survey (DES). Funding for the DES Projects has been provided by the U.S. Department of Energy, the U.S. National Science Foundation, the Ministry of Science and Education of Spain, the Science and Technology Facilities Council of the United Kingdom, the Higher Education Funding Council for England, the National Center for Supercomputing Applications at the University of Illinois at Urbana-Champaign, the Kavli Institute of Cosmological Physics at the University of Chicago, the Center for Cosmology and Astro-Particle Physics at the Ohio State University, the Mitchell Institute for Fundamental Physics and Astronomy at Texas A\&M University, Financiadora de Estudos e Projetos, Funda{\c c}{\~a}o Carlos Chagas Filho de Amparo {\`a} Pesquisa do Estado do Rio de Janeiro, Conselho Nacional de Desenvolvimento Cient{\'i}fico e Tecnol{\'o}gico and the Minist{\'e}rio da Ci{\^e}ncia, Tecnologia e Inova{\c c}{\~a}o, the Deutsche Forschungsgemeinschaft, and the Collaborating Institutions in the Dark Energy Survey.
The Collaborating Institutions are Argonne National Laboratory, the University of California at Santa Cruz, the University of Cambridge, Centro de Investigaciones Energ{\'e}ticas, Medioambientales y Tecnol{\'o}gicas-Madrid, the University of Chicago, University College London, the DES-Brazil Consortium, the University of Edinburgh, the Eidgen{\"o}ssische Technische Hochschule (ETH) Z{\"u}rich,  Fermi National Accelerator Laboratory, the University of Illinois at Urbana-Champaign, the Institut de Ci{\`e}ncies de l'Espai (IEEC/CSIC), the Institut de F{\'i}sica d'Altes Energies, Lawrence Berkeley National Laboratory, the Ludwig-Maximilians Universit{\"a}t M{\"u}nchen and the associated Excellence Cluster Universe, the University of Michigan, the National Optical Astronomy Observatory, the University of Nottingham, The Ohio State University, the OzDES Membership Consortium, the University of Pennsylvania, the University of Portsmouth, SLAC National Accelerator Laboratory, Stanford University, the University of Sussex, and Texas A\&M University.
Based in part on observations at Cerro Tololo Inter-American Observatory, National Optical Astronomy Observatory, which is operated by the Association of Universities for Research in Astronomy (AURA) under a cooperative agreement with the National Science Foundation.}
\end{acknowledgments}

\begin{center}
\textbf{DATA AVAILABILITY}
\end{center}
{
All data is available on \href{https://cosmohub.pic.es}{cosmohub.pic.es} (DES Y3 weak lensing shape catalog, DES Y3 GOLD catalog, Euclid FS2 and MICECAT V2), \href{https://datalab.noirlab.edu}{datalab.noirlab.edu} (DES Y3 GOLD cutouts and BUZZARD simulation), \href{https://www.javiercarron.com/catalogue}{www.javiercarron.com/catalogue} (SDSS DR16 filaments catalog), \href{https://sites.google.com/site/yenchicr/catalogue-data?authuser=0}{sites.google.com/site/yenchicr/catalogue-data} (SDSS DR12 filaments catalog), and \href{https://des.ncsa.illinois.edu/releases/y3a2/gal-morphology}{des.ncsa.illinois.edu/releases/y3a2/gal-morphology} (morphological classifications), 
}

\begin{center}
\textbf{CODE AVAILABILITY}
\end{center}

{
Python code for the estimator is hosted at \href{https://github.com/pdsferreira/laia}{github.com/pdsferreira/laia}.
}

\begin{center}
\textbf{AUTHOR CONTRIBUTIONS}
\end{center}

{
PDSF conceived the study; developed the estimator, methodology, and code; performed the statistical analysis; contributed to the interpretation; drafted the manuscript; and assembled the data products.
ROR co-developed the estimator and code and contributed to interpretation and drafting of the manuscript.
PSF developed the residual spurious correlations removal code, contributed to the statistical analysis and interpretation, performed visual inspection of 10\,000 galaxies and critically revised the manuscript.
AC advised on and assisted with the visual-inspection protocol, contributed to the galaxy-classification pipeline, and helped define the quality-cut criteria.
FF performed \textsc{Morfometryka}-based consistency checks of position angles and World Coordinate System solutions.
VM provided independent methodological scrutiny and robustness checks, contributed to the interpretation, and critically revised the manuscript.
CRB critically revised the manuscript. 
RC advised on the statistical analysis and critically revised the manuscript.
}

\vspace{1.2\baselineskip}
\noindent\rule{\columnwidth}{0.4pt}
\vspace{0.0\baselineskip}

\setcounter{figure}{0}

\makeatletter
\renewcommand{\theHfigure}{ED.\arabic{figure}}

\renewcommand{\fnum@figure}{Supplementary Figure~\thefigure}%
\makeatother

\vspace{3.2cm}
\begin{center}
\textbf{SUPPLEMENTARY INFORMATION}
\end{center}
\vspace{-0.3cm}

\begin{figure}[h]
\includegraphics[width=0.97\linewidth]{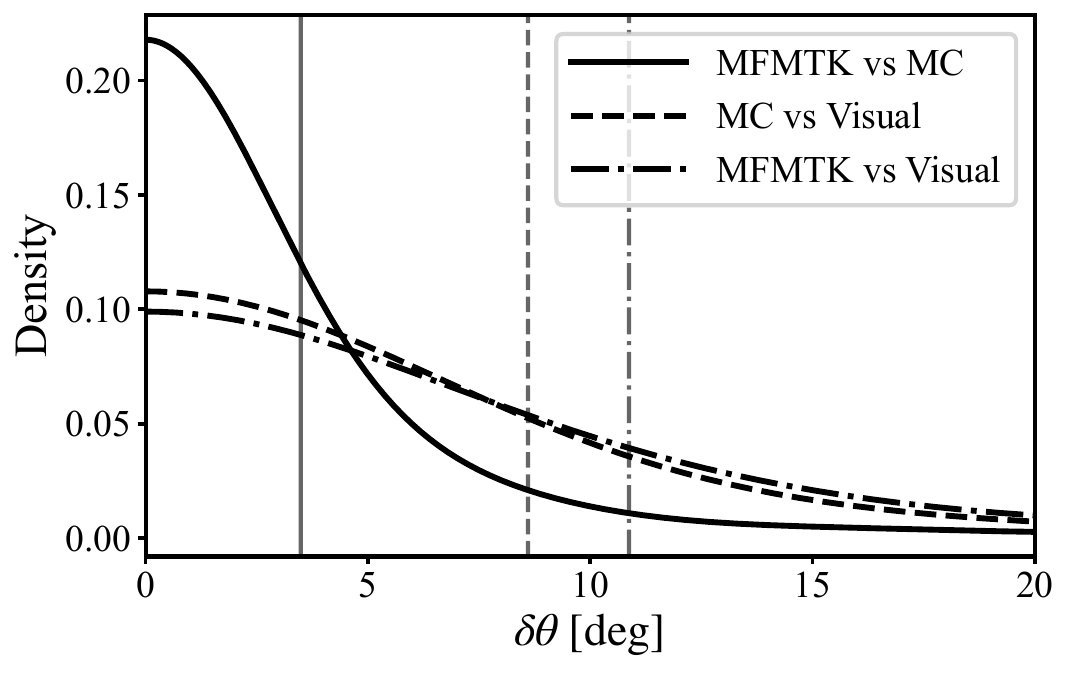}
\caption{\textbf{Position angle consistence across different methods.} Distribution of position-angle differences, $\delta\theta$, for a sample of 10{,}000 galaxies, comparing \textsc{Morfometryka} (MFMTK), \texttt{metacalibration} (MC), and visual inspection. Vertical lines indicate the 68.27\% ($1\sigma$) upper bound for each distribution. MFMTK and visual inspection use only the $I$ band, whereas \texttt{metacalibration} combines the $R$, $I$, and $Z$ bands to estimate the PA.}
\label{fig:delta_theta}
\end{figure}

\begin{figure}[!h]
    \includegraphics[width=0.98\linewidth]{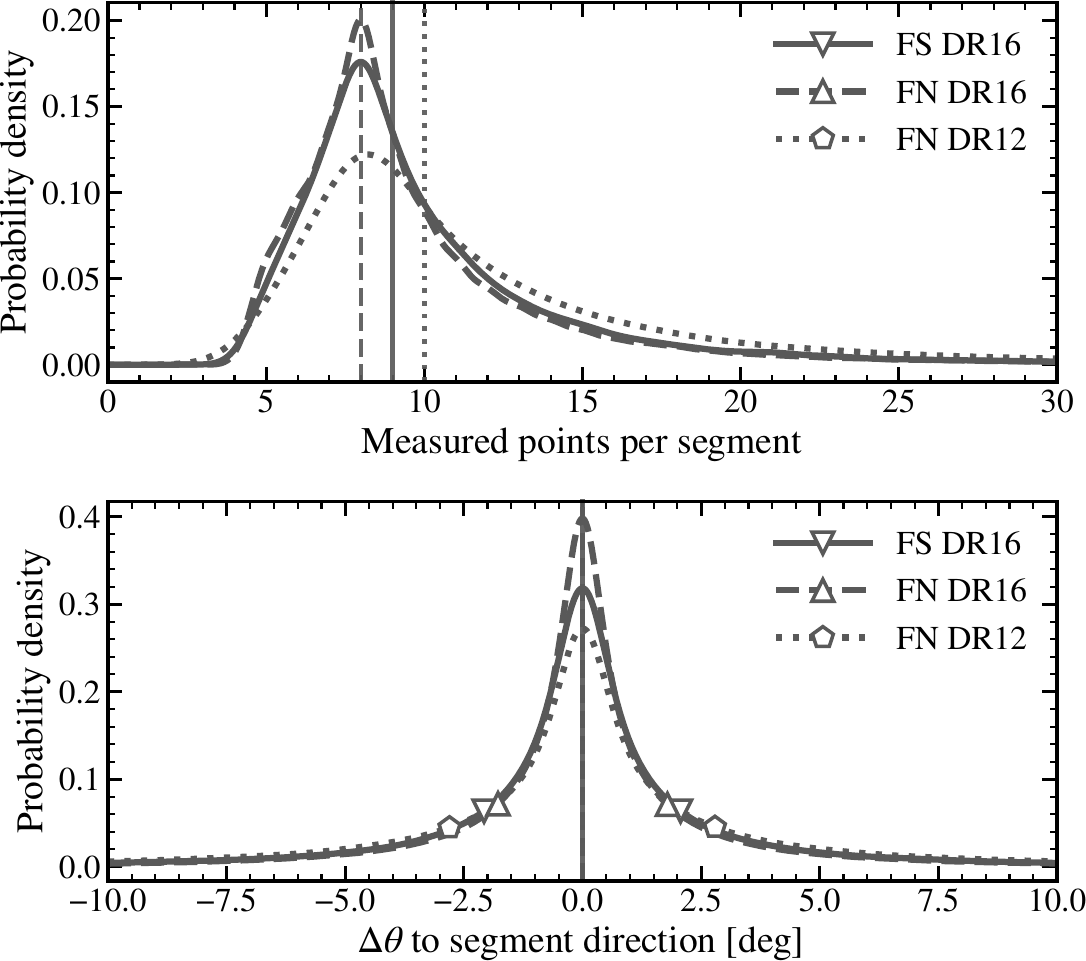}
\caption{\textbf{Sampling and orientation stability of the fixed-length filament segments.}
Kernel-density estimates for the retained segment samples. Left, number of native catalogue spine samples contributing to each fixed $10\,h^{-1}{\rm Mpc}$ segment after requiring at least five contributing spine samples. Right, signed axial residual $\Delta\theta$ between each contributing local spine tangent direction and the final segment PA. Grey curves denote FS DR16, FN DR16 and FN DR12 with solid, dashed and dotted line styles, respectively, and are normalized to unit area. Vertical lines mark medians; in the $\Delta\theta$ panel, downward triangles, upward triangles and pentagons mark the quantile-equivalent $\pm1\sigma$ intervals, defined by the 15.9th and 84.1st percentiles, for FS DR16, FN DR16 and FN DR12. These diagnostics show that the fiducial segment PAs are supported by multiple local spine samples and are stable against small-scale sampling fluctuations.}
    \label{fig:filament_stats}
\end{figure}
\end{document}